\documentclass[aps,prd,twocolumn,superscriptaddress,preprintnumbers,floatfix,nofootinbib,notitlepage,showkeys,showpacs]{revtex4-1}

\usepackage[utf8]{inputenc}

\usepackage[normalem]{ulem}

\usepackage{graphicx}
\usepackage{hyperref}
\usepackage{latexsym}
\usepackage{amsmath}
\usepackage{amssymb}
\usepackage{bbm}
\usepackage{pdfsync}
\usepackage{epsfig}
\usepackage{epstopdf}
\usepackage{subfigure}
\usepackage[dvipsnames]{xcolor}
\usepackage{comment}
\usepackage{slashed}
\usepackage{multirow}

\newcommand\rout{\bgroup\markoverwith{\textcolor{red}{\rule[0.5ex]{2pt}{0.4pt}}}\ULon}

\begin{document}

\title{Strategies to Detect Dark-Matter Decays with Line-Intensity Mapping}
\author{Jos\'e Luis Bernal}
\affiliation{Department of Physics and Astronomy, Johns Hopkins University, 3400 North Charles Street, Baltimore, Maryland 21218, USA}
\author{Andrea Caputo}
\affiliation{School of Physics and Astronomy, Tel-Aviv University, Tel-Aviv 69978, Israel}\affiliation{ Department of Particle Physics and Astrophysics,
Weizmann Institute of Science, Rehovot 7610001,Israel}\affiliation{Max-Planck-Institut f\"ur Physik (Werner-Heisenberg-Institut), F\"ohringer Ring 6, 80805 M\"unchen, Germany}
\author{Marc Kamionkowski}
\affiliation{Department of Physics and Astronomy, Johns Hopkins University, 3400 North Charles Street, Baltimore, Maryland 21218, USA}

\begin{abstract}
The nature of dark matter is a longstanding mystery in cosmology, which can be studied with laboratory or collider experiments, as well as astrophysical and cosmological observations. In this work, we propose realistic and efficient strategies to detect radiative products from dark-matter decays with line-intensity mapping (LIM) experiments. This radiation will behave as a line interloper for the atomic and molecular spectral lines targeted by LIM surveys. The most distinctive signatures of the contribution from dark-matter radiative decays are an extra anisotropy on the LIM power spectrum due to projection effects, as well as a narrowing and a shift towards higher intensities of the voxel intensity distribution.  We forecast the minimum rate of decays into two photons that LIM surveys will be sensitive to as function of the dark-matter mass in the range $\sim 10^{-6}-10$ eV, and discuss how to reinterpret such results for dark matter that decays into a photon and another particle. We find that both the power spectrum and the voxel intensity distribution are expected to be very sensitive to the dark-matter contribution, with the voxel intensity distribution being more promising for most experiments considered. Interpreting our results in terms of the axion, we show that LIM surveys will be extremely competitive to detect its decay products, improving several orders of magnitudes (depending on the mass) the sensitivity of laboratory and astrophysical searches, especially in the mass range $\sim 1-10$ eV.
\end{abstract}

\maketitle

\section{Introduction}
\label{sec:intro}
Cold dark matter, which interacts with baryons only through gravity, is a cornerstone of the successful standard model of cosmology, $\Lambda$CDM. However, there is yet to be a microscopic model for dark matter backed by experimental and observational evidence. There is a vast variety of dark-matter candidates~\cite{Bertone_DMreview}, which can be probed by astrophysical and cosmological observations (see e.g.,~\cite{Gaskings_DMreview} for a recent review), as well as direct detection~\cite{Schumann_DMdirect} and collider~\cite{Boveia_DMcollider} experiments. 

Some of these models involve a very weak coupling with baryons that prompts the decay of dark matter into a photon line; some examples include the axion (a pseudo-Goldstone boson proposed to solve the strong-CP problem that turns out to be a natural dark-matter candidate)~\cite{Abbott:1982af, Dine:1982ah,Preskill:1982cy, Weinberg:1977ma, Wilczek:1977pj, Peccei:1977hh, Peccei:1977ur} and sterile neutrinos~\cite{Kusenko_SterNu}.  There is an ongoing endeavor to study and constrain this possibility, looking for photons produced in dark-matter decays with astrophysical observations at various energies (hence probing different dark-matter masses), such as radio~\cite{Caputo:2018ljp, Caputo:2018vmy}, IR~\cite{Gong:2015hke},  X-rays~\cite{Caputo:2019djj, Boyarsky:2014ska, Riemer-Sorensen:2014yda, Dessert:2018qih}, or gamma-rays~\cite{Blanco_gammaray,Cohen_gammaray}. Short-lived dark-matter particles (with lifetimes $\tau\lesssim 10^{12}$ s) are constrained by spectral distortions~\cite{Ellis_gravitino, Hu_DMSpecD, Chluba_specD} and Big-Bang Nucleosynthesis~\cite{Iocco_BBN,Pospelov_BBN,Poulin_BBNDM}, while CMB anisotropies strongly disfavor lifetimes $\lesssim 10^{25}$ s if all dark-matter decays~\cite{Poulin_decDM,Slatyer_dark,Slatyer_decDM,Lucca_synergy}. 

Line-intensity mapping (LIM) is an emerging observational technique that uses the integrated intensity at a given observed frequency as target observable, holding much promise for both astrophysical and cosmological research~\cite{Kovetz_LIMstatus,Kovetz_astro2020}. Most proposals to characterize dark matter involving LIM observations rely on the effect of the exotic dark matter in the surrounding medium. For instance, exotic energy injection (e.g., from dark-matter decay or annihilation into Standard Model particles, and the subsequent cascades) heats and ionizes the intergalactic medium. This makes the neutral hydrogen hyperfine transition mean intensity and its fluctuations, both depending on the neutral fraction and gas temperature, a powerful probe of dark matter (see e.g.,~\cite{Furlanetto_decay, Shchekinov_21cm, Valdes_dm, Evoli_DM, Munoz_DMscatDA, Liu_DMEoR, Bernal_SMBH, Mena_SMBH, Short_decay}). 

However, given that LIM experiments use the information from all incoming photons, they have the potential to directly observe the electromagnetic radiation produced in dark-matter decays. This way, it is possible to circumvent most of the astrophysical uncertainties that may limit the inference of dark-matter properties from their effect on the 21 cm line intensity.  Following this idea, and inspired by previous works~\cite{Bershady_axion,Gong:2015hke,Grin_axion},  Ref.~\cite{Creque-Sarbinowski_LIMDM} proposed using LIM surveys to search for radiative dark-matter decays in the extragalactic background light. 

In this paper, we build upon this prior work by proposing realistic and feasible strategies to detect the radiation from such decays using LIM surveys, which will target atomic and molecular lines. 
The observed redshift in a LIM survey is derived from the relation between the observed frequency and the rest-frame frequency of the targeted spectral line. However, line emission at different frequencies may redshift into the observed frequency, although its signal will be interpreted as coming from the expected observed redshift anyways. These `line interlopers' contaminate line-intensity maps and need to be cleaned or taken into account to avoid biased cosmological or astrophysical parameter inference (see e.g., Refs.~\cite{Oh2003,Wang2006,Visbal_mask, Liu2011,Breysse_foregrounds, Lidz_interlopers, Sun_foregrounds, Cheng_deconfInterlopers, Gong_interlopers}). In the case of monoenergetic photons from dark-matter decays, there will be an additional spectral line, the frequency of which will be determined by the masses of the dark matter and the other daughter particle. In this scenario, this emission will unexpectedly act as an interloper of the targeted line. We propose to turn this contamination into the signal of interest.

We describe the modeling of the effect of radiative  dark-matter decays as interlopers for the LIM power spectrum and voxel intensity distribution (the distribution of measured intensities in each voxel, i.e., three-dimensional pixels) of atomic and molecular spectral lines. The first feature of a dark-matter line is an increase in the anisotropy of the power spectrum, which is imprinted on the quadrupole and higher multipoles of the clustering. The additional radiation from dark-matter decays also shifts the voxel intensity distribution towards higher intensities, and at the same time narrows it. 

We forecast the sensitivity of current and forthcoming LIM surveys to detect radiative dark-matter decays following these strategies as a function of the dark-matter mass. We focus on the dark-matter decays into two photons, but provide an easy way to reinterpret our results for other radiative dark-matter decays. Our results are very promising, improving several orders of magnitudes the sensitivity of current studies using the CMB power spectrum. Interpreting our results in the framework of axion dark matter, which decays into two photons, we show that LIM surveys will be extremely competitive. LIM experiments have the potential to dramatically improve current and forecasted bounds, especially in the mass range $\sim 1-10$~eV, opening the possibility of a detection of the QCD axion. 

Unlike Ref.~\cite{Creque-Sarbinowski_LIMDM}, our strategies do not rely on cross-correlations with other tracers of the large-scale structure; this thus facilitates detection of decays from higher redshifts where tracers may be lacking. This means that we can probe more masses with the same observed frequency band. We also propose the use of the voxel intensity distribution, which turns out to be more sensitive to the contribution from dark-matter decays than the power spectrum. In addition, following the strategies proposed in this work grants access to features of the dark-matter contribution to LIM observables that depend on the redshift of decay, hence allowing for the determination of the dark-matter mass in case of detection. Furthermore, these strategies are completely achievable and realistic. Although we focus on the power spectrum and the voxel intensity distribution, our proposal can be easily extended to other summary statistics. 

This article is structured as follows. We review the formalism regarding the computation of the LIM observables considered in Sec.~\ref{sec:LIM}; present the model for such observables using the electromagnetic radiation from dark-matter decays as a spectral line in Sec.~\ref{sec:DMDIM}; propose the strategies to detect the intensity of photons from dark-matter decays  in Sec.~\ref{sec:strategies}; report the detection potential sensitivity from ongoing and forthcoming LIM surveys in Sec.~\ref{sec:forecasts}; and discuss the results and conclude in Sec.~\ref{sec:discussion} and Sec.~\ref{sec:conclusions}, respectively. Technical details about the model and calculations are provided in the Appendices. 

Throughout this work we assume a fixed $\Lambda$CDM cosmology (with the exception of the decaying dark matter), taking the best-fit parameter values from \textit{Planck} temperature, polarization and lensing power spectra~\cite{Planck18_parameters} as our fiducial values: a reduced Hubble constant $h=0.6736$, physical dark-matter and baryon densities today $\Omega_{\rm dm}h^2 = 0.12$ and $\Omega_{\rm b} = 0.02237$, respectively, a spectral index $n_s = 0.9649$ and an amplitude $A_s=2.1 \times 10^{-9}$ of the primordial scalar power spectrum.

\section{LIM observables}
\label{sec:LIM}

The brightness temperature $T$ observed for a given line in a LIM experiment is related to the line's rest-frame frequency $\nu$ and emission redshift $z$, by\footnote{We use the terms `temperature' (by which we mean brightness temperature) and `intensity' interchangeably hereinafter; the relation between the two is spelled out in App.~\ref{app:LIMPk}.}
\begin{equation}
     T(z) = \frac{c^3(1+z)^2}{8\pi k_{\rm B}\nu^3H(z)}\rho_{\rm L}(z) = X_{\rm LT}(z)\rho_{\rm L}(z)\,,
\end{equation}
where $c$ is the speed of light, $k_{\rm B}$ is the Boltzmann constant and $H$ is the Hubble expansion rate~\cite{Lidz_2011}. The luminosity density $\rho_L(z)$ in the line depends on the spectral line under consideration, but in general (except for the HI line during and before reionization) it is assumed that all radiation comes from halos.\footnote{The Lyman-$\alpha$ line is an ambiguous case, since the observed emission may extend beyond the dark-matter halo due to radiative transfer. Nonetheless, we assume that this is not the case and leave the study of such a scenario for future work.} Therefore, in terms of the mass function $(dn/dM)(M,z)$ and mass-luminosity relation $L(M,z)$, it is,
\begin{equation}
    \langle \rho_{\rm L}\rangle(z) = \int {\rm d}ML(M,z)\frac{{\rm d}n}{{\rm d}M}(M,z).
\label{eq:rhoL}
\end{equation}

LIM fluctuations combine astrophysical and cosmological dependence and are very non-Gaussian, so that several summary statistics have been proposed to exploit the information contained in them. In this work we focus on the power spectrum and the voxel intensity distribution (VID). We use \texttt{lim}\footnote{\url{https://github.com/jl-bernal/lim}} to compute all the LIM observables and related quantities.

\subsection{LIM power spectrum}
\label{sec:LIMPk}
In this subsection we describe our modeling of the LIM power spectrum and its covariance, following Ref.~\cite{Bernal_IM}. We refer the reader to that reference and to App.~\ref{app:LIMPk} for more details. 

Since the emission originates in halos, the temperature fluctuation is a biased tracer of the matter density fluctuations. In addition, there is a scale-independent shot-noise contribution to the LIM power spectrum due to the discrete distribution of halos. Adding the clustering and the shot-noise contributions, the total anisotropic LIM power spectrum is given by
\begin{equation}
    \begin{split}
        P(k,\mu, z) & = \langle T\rangle^2(z)b^2(z)F_{\rm rsd}^2(k,\mu,z)P_{\rm m}(k,z) + \\
        & + X_{\rm LT}^2\int{\rm d}ML^2(M,z)\frac{{\rm d}n}{{\rm d}M}, 
    \end{split}
\end{equation} 
where $\mu \equiv \boldsymbol{k}\cdot\boldsymbol{k}_\parallel/k^2$ is the cosine of the angle between the Fourier mode $\boldsymbol{k}$ and its component $\boldsymbol{k}_\parallel$ along the line of sight, $b(z)$ is the luminosity-weighted linear halo bias, $F_{\rm rsd}$ is a factor encoding the effect of redshift-space distortions, and $P_{\rm m}(k,z)$ is the matter power spectrum. Regarding $F_{\rm rsd}$, we include the Kaiser effect at large scales~\cite{Kaiser1987} and a suppression at small scales, determined by the velocity dispersion $\sigma_{\rm v}$, for which we take the linear prediction, $\sigma_{\rm v}^2(z)=\int P_{\rm m}(k,z){\rm dk}/6\pi^2$, as the fiducial value. 
LIM observations are limited by the spectral and angular resolutions, and also by the volume probed by the experiment or the presence of foregrounds. Therefore, there are modes that are inaccessible. This can be modeled applying window functions to the power spectrum: $W_{\rm res}$ to model the smoothing due to instrumental resolution, and $W_{\rm vol}$ to account for the finite size of the volume probed and potential presence of foregrounds. Thus, the measured LIM power spectrum is given by
\begin{equation}
    \tilde{P}(k,\mu,z) =   W_{\rm vol}(k,\mu,z)W_{\rm res}(k,\mu,z)P(k,\mu,z).
\label{eq:obsPk}
\end{equation}

It is not possible to obtain a well defined $\mu$ from actual observations due to the change of the line of sight with the pointing. Nonetheless, using e.g., the Yamamoto estimator~\cite{YamamotoEstimator}, it is possible to directly measure the Legendre multipoles of the LIM power spectrum, given by
\begin{equation}
\begin{split}
    \tilde{P}_{\ell}(k) = \frac{2\ell+1}{2}  \int_{-1}^{1}{\rm d}\mu \tilde{P}(k\mu)\mathcal{L}_{\ell}(\mu),
\end{split}
\label{eq:multipole_scale}
\end{equation}
where $\mathcal{L}_{\ell}$ is the Legendre polynomial of degree $\ell$. 

The limitations of LIM experiments also introduce instrumental white noise in the measurement. The variance of such noise depends on whether the signal from each antenna is cross-correlated with the rest or not, and is given by (see e.g., Ref.~\cite{Bull_21cm})
 \begin{equation}
    P_{\rm N}^{\rm dish}  = \frac{\sigma_{\rm N}^2V_{\rm vox}}{N_{\rm ant}}, \qquad P_{\rm N}^{\rm interf} = \frac{\sigma_{\rm N}^2V_{\rm vox}\Omega_{\rm FoV}}{n_{\rm s}},
\label{eq:Pn}
 \end{equation}
 where $\sigma_{\rm N}^2$ is the instrumental noise variance per voxel per antenna, $V_{\rm vox}$ is the volume of the voxel (the size of which is assumed here to be determined by the instrumental resolution unless otherwise stated), $\Omega_{\rm FoV}=c^2/\left(\nu_{\rm obs}D_{\rm dish}\right)^2$ is the field of view, $\nu_{\rm obs}$ is the observed frequency, $D_{\rm dish}$ is the diameter of each antenna, and $n_{\rm s}$ is the average number density of baselines. 

Assuming Gaussianity, neglecting mode coupling, and accounting for the instrumental noise and sample variance, the variance per $k$ and $\mu$ bins is given by $\tilde{\sigma}^2(k,\mu)\equiv [\tilde{P}(k,\mu)+P_{\rm N}]^2/N_{\rm modes}$, where $N_{\rm modes}$ is the number of modes observed per bin. The total covariance matrix for the multipoles of the LIM power spectrum is the combination of the subcovariance matrices of each multipoles and those between different multipoles. The subcovariance matrix between the multipoles $\ell$ and $\ell^\prime$ is given by
\begin{equation}    
    \begin{split}
        \tilde{C}_{\ell\ell^\prime}(k) & = \frac{\left(2\ell +1\right)\left(2\ell^\prime +1 \right)}{2}\times \\
        & \times \int_{-1}^{1}{\rm d}\mu \tilde{\sigma}^2(k,\mu)\mathcal{L}_\ell(\mu)\mathcal{L}_{\ell^\prime}(\mu).
    \end{split}
    \label{eq:covariance}
\end{equation}

\subsection{Voxel intensity distribution}
\label{sec:LIMVID}
Exploiting the 1-point distribution function of the brightness temperature allows to access the non-Gaussian components of its fluctuations, that are not included in the power spectrum. Moreover, it has been shown that, while the LIM power spectrum can efficiently constrain cosmology, the VID is a very powerful summary statistic to constrain the luminosity function of the spectral line of interest~\cite{Breysse_VID}. We follow the formalism described in Ref.~\cite{Breysse_VID}, which is briefly reviewed below.

The VID is the probability distribution function $\mathcal{P}(T)$ for the measured brightness temperature in a voxel, which is related with the luminosity function and the number $N_{\rm e}$ of halos or emitters in the voxel. Observing a voxel that contains no emitter with a perfect experiment (i.e., without noise) returns zero brightness temperature, hence $\mathcal{P}_0=\delta_D(T)$, where $\delta_D$ is the Dirac delta. In turn, the brightness temperature observed in a given voxel containing $N_{\rm e}$ emitters is:
\begin{equation}
    T = \frac{X_{\rm LT}}{V_{\rm vox}}\sum_i^{N{_{\rm e}}} L_i,
\label{eq:T_voxel}    
\end{equation}
where $L_i$ is the luminosity of each emitter~\cite{Lidz_2011}. We neglect edge and beam effects, which is equivalent to consider that a source contributes entirely to the voxel in which it is contained.

The probability of  observing a brightness temperature $T$ in the voxels that only contain one emitter is:
\begin{equation}
    \mathcal{P}_1(T) = \frac{V_{\rm vox}}{\bar{n}X_{\rm LT}}\left.\frac{{\rm d} n}{{\rm d} L}	\right\lvert_{L=\rho_L(T)V_{\rm vox}}\, ,
\label{eq:probT1}
\end{equation}
where $\bar{n}$ is the mean comoving number density of emitters. Now consider two emitters $a$ and $b$; the sum of their emission is $T_a+T_b=T$. Accounting for the probability distribution function of each emitter:
\begin{equation}
    \mathcal{P}_2(T) = \int {\rm d}T_a \mathcal{P}_1(T_a)\mathcal{P}_1(T-T_a) = \left(\mathcal{P}_{1}*\mathcal{P}_1\right)(T)\,,
    \label{eq:ProbT2}
\end{equation}
where `$*$' denotes the convolution operator. Repeating this argument iteratively, the probability of observing a brightness temperature $T$ in a voxel with $N_{\rm e}$ emitters is
\begin{equation}
    \mathcal{P}_{N_{\rm e}}(T)=\left(\mathcal{P}_{N_{\rm e}-1}*\mathcal{P}_1\right)(T)\,.
\label{eq:probTN}
\end{equation}
The conditional probability of a voxel with total brightness temperature $T$ containing $N_{\rm e}$ is given by the product of $\mathcal{P}_{N_{\rm e}}(T)$ and the probability $\mathcal{P}_{\rm e}(N_{\rm e})$ of the voxel containing such number of emitters. Therefore, the total probability density distribution of the brightness temperature in a voxel is the sum of the conditional probabilities mentioned above:
\begin{equation}
    \mathcal{P}_{\rm astro}(T)=\sum_{N_{\rm e}=0}^\infty \mathcal{P}_{N_{\rm e}}(T)\mathcal{P}_{\rm e}(N_{\rm e}).
\label{eq:Prob_Tsignal}
\end{equation}

However, there is no perfect experiment and there will always be some instrumental noise which contributes to the total brightness temperature measured in the voxel. We assume that the instrumental noise per voxel follows Gaussian distribution with zero mean and variance $P_{\rm N}/V_{\rm vox}$ (i.e., the final survey instrumental variance per voxel). Therefore, the probability of the instrumental noise being $T$ in a voxel is
\begin{equation}
    \mathcal{P}_{\rm noise}(T) = \sqrt{\frac{V_{\rm vox}}{2\pi P_{\rm N}}}\exp\left\lbrace -\frac{T^2V_{\rm vox}}{2P_{\rm N}}\right\rbrace\,.
    \label{eq:PDFnoise}
\end{equation}
Similarly to the contribution from the cases with different number of emitters, the contribution from instrumental noise is added to the VID convoluting Eqs.~\eqref{eq:Prob_Tsignal} and~\eqref{eq:PDFnoise}:
\begin{equation}
    \mathcal{P}_{\rm tot} = \left(\mathcal{P}_{\rm signal} * \mathcal{P}_{\rm noise}\right)(T)\,.
    \label{eq:VIDtot}
\end{equation}
In practice, continuous probability distribution functions are impossible to obtain directly from intensity maps. Instead, the VID can be inferred through the computation of histograms of the number $B_i$ of voxels for which the measured $T$ is within a given brightness temperature bin with width $\Delta T_i$:
\begin{equation}
    B_i = N_{\rm vox}\int_{T_i}^{T_i+\Delta T_i} \mathcal{P}_{\rm tot}(T){\rm d}T\,,
    \label{eq:VIDhist}
\end{equation}
where $N_{\rm vox}$ is the number of voxels that the volume probed is divided into. We assume that the bins follow a Poisson distribution, hence the variance $\sigma_i^2$ of $B_i$ is equal to $B_i$. This is a good approximation for bins containing many pixels, which in the end are the ones that dominate the signal-to-noise ratio. Finally, although each voxel is located at slightly different redshift, we compute Eq.~\eqref{eq:VIDhist} at the mean redshift of the volume probed (as it is customary for the power spectrum, too).

The combination of the VID and the power spectrum was explored in Ref.~\cite{Ihle_VID-PS}, accounting for their covariance using simulations. We leave the combination of both summary statistics for future work, and consider them independently.

\section{Dark-matter decay intensity mapping}
\label{sec:DMDIM}
We focus this section and the results shown hereafter on dark matter that decays into two photons. We discuss other radiative decay channels and the corresponding reinterpretation of our results at the end of the section.  We limit our analysis to radiative decays of dark matter because LIM techniques require a narrow, well-defined emission line to target. 

Consider a dark-matter model in which  a fraction $f_{\rm \chi}$ of all of the dark matter is made of particles $\chi$ with mass $m_\chi$. While we assume that the rest of the dark matter is standard cold dark matter, $\chi$ particles decay into Standard Model particles, with a branching ratio $f_{\gamma\gamma}$ for decays into two photons. These photons have a specific energy, depending on $m_{\chi}$, corresponding to a rest-frame frequency
\begin{equation}
    \chi\rightarrow \gamma\gamma;\qquad \nu_{\gamma\gamma}=\frac{m_\chi c^2}{4\pi\hbar}\,,
    \label{eq:nuDM}
\end{equation}
where $\hbar$ is the Planck constant over $2\pi$.  

The mean energy injection by unit of time $t$ and volume $V$ for dark-matter decays into photons is given by~\cite{Chen:2003gz, Finkbeiner:2011dx, Pierpaoli:2003rz}
\begin{equation}
    \left.\frac{\rm d E}{\rm d V\rm d t}\right\lvert_{\rm inj,\gamma\gamma}  = f_{\gamma\gamma} f_\chi \Omega_{\rm dm}\rho_c (1+z)^3 c^2\Gamma_\chi \left(1 + 2 \mathcal{F}_{\gamma}\right) \, ,
\label{eq:Einj}
\end{equation}
where $\Omega_{\rm dm}$ and $\rho_c$ are the total dark-matter density parameter and the critical density of the Universe today, respectively, and $\Gamma_\chi\equiv\tau_\chi^{-1} $ is the decay rate of dark matter into photons, defined as the inverse of the lifetime $\tau_\chi$. We assume that the decay rate is constant in time, and low enough to neglect the reduction in the abundance of dark matter as the Universe evolves. 

The additional term $2\mathcal{F}_\gamma$ in Eq.~\eqref{eq:Einj} accounts for the stimulated decays due to the cosmic microwave background radiation with phase-space distribution $\mathcal{F_\gamma}$~\cite{Caputo:2018ljp,Caputo:2018vmy}. Our results can then be considered conservative, as any other background source, such as the extragalactic background, would enhance the signal. Background radiation may equally induce the inverse decay, but this effect is negligible for the masses and redshifts of interest. While stimulated emission can be significant for low frequencies, it has no effect on our results at $m_{\chi}c^2 \gtrsim 10^{-3}$ eV.

Equation~\eqref{eq:Einj} addresses the energy density rate produced in the form of photons from dark-matter decays, but not all that energy reaches us: some photons interact with matter in their path towards us. Only a fraction $f_{\rm esc}$ of that radiation escapes its environment and reach us. $f_{\rm esc}$ is very small for photons with energies above the excitation energy of hydrogen atoms, 10.2 eV. Therefore, we limit our study to $m_\chi c^2 \lesssim 20\, {\rm eV}$, which corresponds to frequencies $\nu_\gamma \lesssim 4.6\times 10^6$ GHz.

Taking all this into account, the mean luminosity density of radiation created in dark-matter decays is
\begin{equation}
    \langle \rho_{\rm L}^\chi\rangle  = \Theta_\chi \Omega_{\rm dm}\rho_c(1+z)^3c^2(1+2\mathcal{F}_\gamma)\,,
    \label{eq:rhoL_DM}
\end{equation}
where we have defined $\Theta_\chi\equiv f_\chi f_{\gamma\gamma} f_{\rm esc}\Gamma_\chi$ to encode all degenerate dark-matter quantities related with the observed intensity of dark-matter decays.

Note that the luminosity density of dark-matter decays is proportional to the dark-matter density. Therefore, dark-matter decays trace perfectly dark-matter perturbations, which can be approximated as using a unity bias. Moreover, since the distribution of dark matter is not discrete but continuous, there is no shot noise contribution to the power spectrum. Therefore, the dark-matter decay intensity mapping power spectrum is
\begin{equation}
    P_\chi(k,\mu,z) = X_{\rm LT}^2\langle \rho_{\rm L}^\chi\rangle^2(z) F^2_{\rm rsd}(k,\mu,z)P_{\rm m}(k,z)\,.
    \label{eq:DMPK}
\end{equation}
Similarly, the VID associated to dark-matter decays is related with the probability distribution function $\mathcal{P}_{\breve{\rho}}$ of the normalized total matter density $\breve{\rho}_{\rm m}\equiv \rho_{\rm m}/\bar{\rho}_{\rm m}$, where $\bar{\rho}_{\rm m}$ is the mean matter density, as
\begin{equation}
    \mathcal{P}_\chi(T) = \frac{\mathcal{P}_{\breve{\rho}}(\breve{\rho}_{\rm m})}{X_{\rm LT}\Theta_\chi\Omega_{\rm dm}\rho_c(1+z)^3c^2(1+2\mathcal{F}_\gamma)}\, .
    \label{eq:PDFchi}
\end{equation}
$\mathcal{P}_{\breve{\rho}}$ depends on the variance $\sigma_{\rm m}^2$ of the matter perturbations smoothed over a voxel. We use different models for $\mathcal{P}_{\breve{\rho}}$ depending on redshift and value of $\sigma_{\rm m}$. For $z<1$ and $\sigma_{\rm m}>0.7$ we use a double-exponential probability distribution function~\cite{Klypin_PDF}
\begin{equation}
    \mathcal{P}_{\breve{\rho},{\rm DE}}(\breve{\rho}_{\rm m}) = \mathcal{A}\breve{\rho}_{\rm m}^\varsigma\exp\left\lbrace -\left(\frac{\rho_0}{\breve{\rho}_{\rm m}}\right)^{1.1} -\left(\frac{\breve{\rho}_{\rm m}}{\rho_1}\right)^{0.55}\right\rbrace\,,
    \label{eq:doublePDF}
\end{equation}
which has been shown to provide a good fit to simulations. We use fitting functions for the parameters $\varsigma\approx -2$, $\rho_0$ and $\rho_1$ that can be found in App.~\ref{app:PDFchi}, while $\mathcal{A}$ is a normalization factor~\cite{Klypin_PDF}. For higher redshifts or lower values of $\sigma_{\rm m}$ we assume the more general lognormal distribution, which has been for long proposed as a phenomenological fit to the total matter distribution for both observations and
N-body simulations~\cite{Coles_LogNormal, Kayo_PDF01, Kofman:1993mx, Hurtado-Gil:2017dbm}:
\begin{equation}
    \mathcal{P}_{\breve{\rho}_,{\rm LN}}(\breve{\rho}_{\rm m}) = \frac{\exp\left\lbrace -\frac{\left(\log\breve{\rho}_{\rm m}+\sigma_{\rm LN}^2/2\right)^2}{2\sigma^2_{\rm LN}}\right\rbrace}{\breve{\rho}_{\rm m}\sqrt{2\pi\sigma_{\rm LN}^2}}\, ,
    \label{eq:LNPDF}
\end{equation}
where $\sigma_{\rm LN}^2\equiv \log(1+\sigma_{\rm m}^2)$.

We also notice that there has been a significant effort to calculate the probability distribution function of matter fluctuations from first principles, especially using a
path-integral approach~\cite{Bernardeau_2002, Valageas_2002, Valageas:2001td, Ivanov:2018lcg}. In particular, Ref.~\cite{Ivanov:2018lcg} highlights that the lognormal probability distribution function leads to results that are very similar to those theoretically derived. We leave an exploration of the systematic uncertainty associated to the choice of functional form for $\mathcal{P}_{\breve{\rho}}$ for future work, and  limit our analysis to the double exponential and lognormal phenomenological models. 

\subsection{Decays into a photon and another particle}
What has been discussed in this section (and the results reported in Sec.~\ref{sec:forecasts}) so far applies to bosonic dark matter that decays into two photons. However, it is also possible for the dark matter (both bosonic and fermionic) to decay into a photon and a daughter particle $\xi$ with mass $m_\xi$. In this case, the photon rest-frame frequency is given by
\begin{equation}
     \chi\rightarrow \gamma\xi;\qquad \nu_\gamma=\frac{m_\chi c^2}{4\pi\hbar}\left(1-\frac{m_\xi^2}{m_\chi^2}\right)\,,
\end{equation}
and the corresponding mean energy injected by unit of time and volume in form of photons through this decay channel is
\begin{equation}
\begin{split}
    \left.\frac{\rm d E}{\rm d V\rm d t}\right\lvert_{\rm inj,\gamma\xi}  & = f_{\gamma\xi} f_\chi \Omega_{\rm dm}\rho_c (1+z)^3 c^2\times \\
    & \times \Gamma_\chi\frac{1-m_\xi^2/m_\chi^2}{2} \, ,
\end{split}
\end{equation}
where in this case $f_{\gamma\xi}$ is the branching ratio of the $\chi$ particles decaying into a photon and a $\xi$ particle. Note that in this case there is no stimulated or inverse decay from the partner particle, since the occupation number for $\xi$ particles is expected to be small, especially if $\xi$ is a fermion. 

All our results for the $\chi\rightarrow \gamma\gamma$ can be reinterpreted through an analogy with the case discussed in this subsection. The analogy is based in a rescaling of the decaying dark-matter mass as
\begin{equation}
    m_\chi^\prime c^2\rightarrow \frac{m_\chi c^2}{1-m_\xi^2/m_\chi^2}\,,
    \label{eq:mass_rescale}
\end{equation}
and a redefinition of the effective parameter:
\begin{equation}
    \Theta_\chi^\prime \rightarrow \Theta_\chi \frac{f_{\gamma\xi}(1-m_\xi^2/m_\chi^2)}{2f_{\gamma\gamma}(1+2\mathcal{F}_\gamma)}\,,
    \label{eq:Theta_rescale}
\end{equation}
where we have denoted the quantities for the $\gamma\gamma$ case (those reported in this work) with a prime. We will report results assuming the decay into two photons without including the stimulation of such decay (i.e., removing the $(1+2\mathcal{F}_\gamma)$ term) to ease the reinterpretation of our analysis. 

Note that there are stringent lower limits for the mass of fermionic particles that are all the dark matter~\cite{Irsic_wDM, Gilman_wDMlensing, Enzi_wDM}, so that the value of $f_\chi$ may be forced to be very small, especially if $m_\xi$ is small. 

\section{Detecting dark-matter decay as LIM interloper}
\label{sec:strategies}
\begin{figure}[t]
 \begin{centering}
\includegraphics[width=\columnwidth]{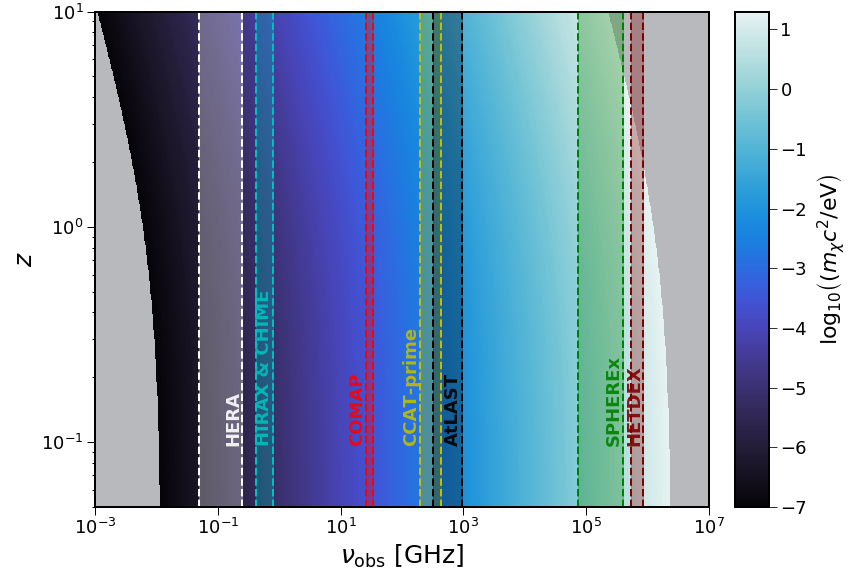}
\caption{Redshift-observed frequency relation for radiative dark-matter decays for different dark-matter masses, assuming massless daughter particles. Frequency bands covered by the LIM experiments considered in this work are marked by colored vertical bands.}  
\label{fig:landscape}
\end{centering}
\end{figure}

There is a plethora of LIM experiments that have from HI to Lyman-$\alpha$ as their main targeted spectral lines, covering all together over seven orders of magnitude in observed frequency.\footnote{See e.g., \url{https://lambda.gsfc.nasa.gov/product/expt/lim_experiments.cfm}} This allows us to probe more than at least eight orders of magnitude of the mass of the decaying dark-matter particles while still fulfilling $m_\chi \lesssim 20\, {\rm eV}/c^2$ so that the produced photons can reach us (in the case of massless daughter particles). We illustrate this scenario with some LIM experiments in Fig.~\ref{fig:landscape}.

If dark matter indeed decays into photons, the radiation emitted would appear as interlopers of LIM experiments that target other spectral lines. We take inspiration from strategies to detect, model and clean atomic or molecular interlopers in LIM experiments (see e.g.,~\cite{Oh2003,Wang2006,Liu2011,Breysse_foregrounds, Lidz_interlopers, Sun_foregrounds, Cheng_deconfInterlopers, Gong_interlopers}) to propose novel ways to probe dark matter with LIM experiments.

In actual LIM observations, there are several line interlopers. Since these lines are known and there is an ongoing program to identify them and model them or remove them from the maps, we do not include them in this work; extending the ideas presented in this article adding known line interlopes is straightforward. Therefore, we consider dark-matter decays as the only interloper of the targeted line with rest-frame frequency $\nu_{\rm l}$. The corresponding redshifts at which the targeted and interloper photons originate are
 \begin{equation}
    z_{\rm l} = \frac{\nu_{\rm l}}{\nu_{\rm obs}}-1,\quad\qquad z_\chi = \frac{\nu_{\gamma\gamma}}{\nu_{\rm obs}}-1\,.
 \label{eq:redshift_of_nu}
 \end{equation}
 
\begin{figure*}
 \begin{centering}
\includegraphics[width=\textwidth]{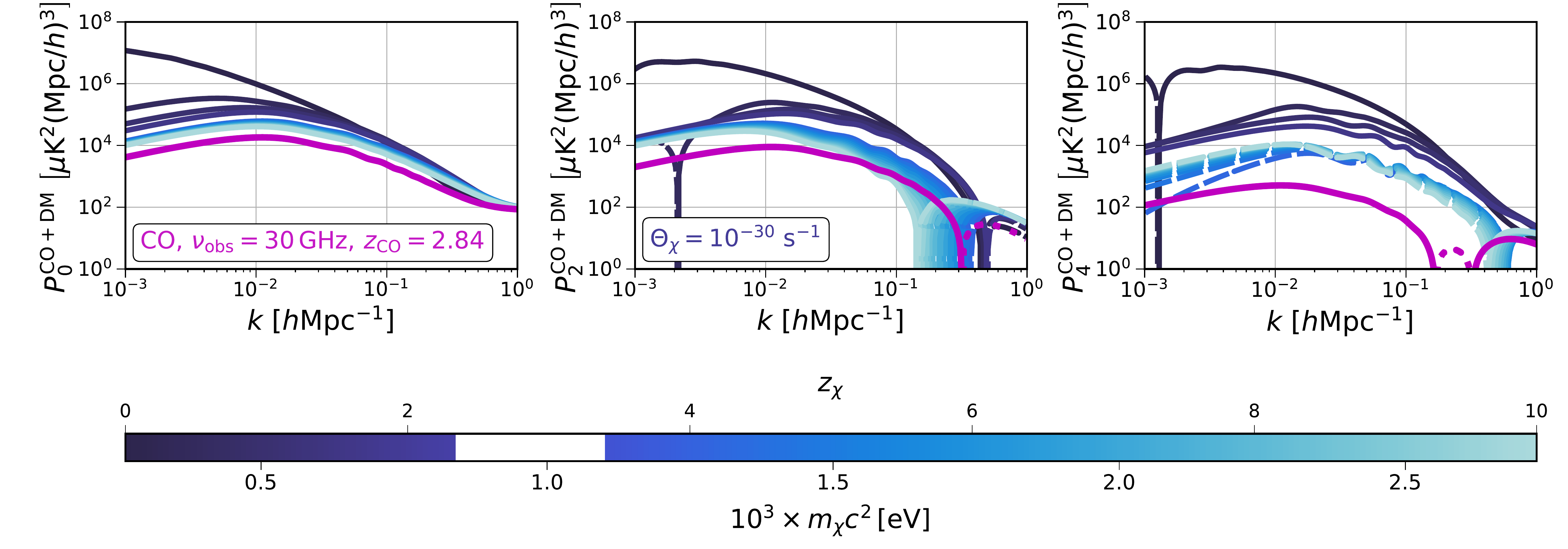}
\caption{Monopole (left), quadrupole (middle) and hexadecapole (right) of the CO power spectrum observed at $\nu_{\rm obs}=30$ GHz (magenta) and the total power spectrum adding the contribution of the projected dark-matter decay power spectrum with $\Theta_\chi=10^{-30}\, {\rm s}^{-1}$. We consider dark-matter masses, corresponding to $z_\chi\in [0,10]$, except for the redshift range corresponding to the volume probed by the CO line (marked in white in the colorbar). Dashed and dotted lines denote negative values. No window functions are applied.}  
\label{fig:Pkproj}
\end{centering}
\end{figure*}

We focus on the effect of line interlopers in the power spectrum, detecting additional contributions  with projection effects~\cite{Lidz_APforeground,Cheng_foregroundsAP,Gong_interlopers}, and in the VID, through the changes in its shape due to additional radiation sources~\cite{Breysse_VID}.

\subsection{Projection effects in power spectrum}
\label{sec:proyection}
In order to compute the LIM power spectrum, redshifts must be converted into distances, and an incorrect redshift estimation introduces projection effects in the measurement. This is what happens when two spectral lines are confused, and it can be used to model and detect the contribution from interlopers. This effect is similar to the Alcock-Paczysnki effect~\cite{Alcock_Paczynski}, for which the anisotropic distortion comes from the difference between the fiducial and the actual cosmology. 

As with the Alcock-Paczynski effect, projection effects can be modeled with rescaling parameters. When dark-matter lines at $z_\chi$ are misidentified as corresponding to the atomic or molecular targeted spectral line coming from $z_{\rm l}$, the inferred distances along the transverse and line-of-sight directions in the volume where the dark-matter decays occur are incorrect by a factor $q_\perp$ and $q_\parallel$, respectively, given by
\begin{equation}
    q_\perp=\frac{D_M(z_\chi)}{D_M(z_l)},\qquad q_\parallel = \frac{(1+z_\chi)/H(z_\chi)}{(1+z_l)/H(z_l)}\,,
\label{eq:scaling}
\end{equation}
where $D_M$ is the comoving angular diameter distance. The correction between inferred and true Fourier modes is $k^{\rm true}_\perp=k^{\rm infer}_\perp/q_\perp$ and $k^{\rm true}_\parallel=k^{\rm infer}_\parallel/q_\parallel$, or, in terms of $k$ and $\mu$~\citep{Ballinger_scaling96}:
\begin{equation}
\begin{split}
    & k^{\rm true} = \frac{k^{\rm infer}}{q_\perp}\left[1+\left(\mu^{\rm infer}\right)^2\left(F_{\rm proj}^{-2}-1\right)	\right]^{1/2}, \\
    & \mu^{\rm true} = \frac{\mu^{\rm infer}}{F_{\rm proj}}\left[1+\left(\mu^{\rm infer}\right)^2\left(F_{\rm proj}^{-2}-1\right)	\right]^{-1/2},
\end{split}
\label{eq:scaling_kmu}
\end{equation}
where $F_{\rm proj} \!\equiv\! q_\parallel/q_\perp$.

Taking this into account, the measured interloper-line power spectrum will be distorted due to those projection effects. In practice, the power spectrum must be computed at $z_\chi$ following Eq.~\eqref{eq:DMPK}, applying the window functions from Eq.~\eqref{eq:obsPk} also at $z_\chi$, and later applying the distortion of scales from Eq.~\eqref{eq:scaling_kmu}. Therefore, the measured multipoles of the projected LIM power spectrum from dark-matter decays is 
\begin{equation}
\begin{split}
    \tilde{P}_{\ell}^\chi&(k^{\rm infer}, z_\chi) =  \frac{1}{q_\perp^2q_\parallel}\frac{2\ell+1}{2}  \\ 
    & \times \int_{-1}^{1}{\rm d}\mu^{\rm infer} \tilde{P}_\chi(k^{\rm true},\mu^{\rm true},z_\chi)\mathcal{L}_{\ell}(\mu^{\rm infer}),
\end{split}
\label{eq:Pmul_rescale}
\end{equation}
where the factor $q_\perp^2q_\parallel$ is due to the isotropic dilation of the volume from the projection effect. The total LIM power spectrum multipoles in the presence of dark-matter decays as interloper is given by 
\begin{equation}
    \tilde{P}_{\ell}^{\rm tot}(k^{\rm infer}) = \tilde{P}_\ell^{\rm l}(k^{\rm infer},z_l)+\tilde{P}^\chi_{\ell}(k^{\rm infer},z_\chi)\,,
\label{eq:Pelltot}
\end{equation}
where we assume that $\nu_{\rm l}$ and $\nu_\chi$ are different enough so that the volumes they come from do not overlap, and therefore there is no cross-correlation between them. Otherwise, Eq.~\eqref{eq:Pelltot} should account for that contribution, but the projection effect would be minimal and the contribution from dark-matter decays would be almost completely degenerate with the clustering power spectrum of the atomic or molecular line. 

Equation~\eqref{eq:covariance} can be used to compute the covariance evaluated at $z_{\rm l}$, but using $\tilde{P}_{\ell}^{\rm tot}$ in this case, since the isotropic dilation from the projection effects cancel the changes in the number of modes depending on the volume probed, and the instrumental noise is projected also in the targeted volume at $z_{\rm l}$.

We illustrate how the LIM power spectrum changes with the addition of the projected contribution from dark-matter decays in Fig.~\ref{fig:Pkproj}. Note that the monopole is completely dominated by the contribution from dark-matter decays for light masses (which means $z_\chi < z_{\rm l}$). However, the large anisotropy introduced by the projection effects significantly modify the quadrupole and hexadecapole of the power spectrum. The level of anisotropy then provides information about the dark-matter mass.  Moreover, exploiting the information in the power spectrum anisotropies allows us to constrain higher dark-matter masses for which $z_\chi>z_{\rm l}$.

\subsection{Shape of the VID}
\label{sec:VIDshape}
As discussed in Sec.~\ref{sec:LIMVID}, any radiation contributes to the temperature measured in a voxel, and its probability distribution function can be included in the total VID convolving it with the rest. In this case, the total brightness temperature is given by $T = T_{\rm l}+T_\chi+T_{\rm noise}$, contributions that follow the probability distribution functions given in Eqs.~\eqref{eq:Prob_Tsignal}, \eqref{eq:PDFchi} and \eqref{eq:PDFnoise}, respectively. Note that $\mathcal{P}_{\rm l}$ and $\mathcal{P}_\chi$ are computed at their respective redshifts. Therefore, the VID for the total brightness temperature is
\begin{equation}
    \mathcal{P}_{\rm tot+\chi}=\left(\left(\mathcal{P}_{\rm l}*\mathcal{P}_\chi\right)*P_{\rm noise}\right)(T).
\label{eq:Prob_InterLine}
\end{equation}

Note that the changes in the VID due to the presence of dark-matter decays is not as straightforward as in the case of the power spectrum. According to Eqs.~\eqref{eq:rhoL_DM} and~\eqref{eq:PDFchi}, at a given $\nu_{\rm obs}$, $T_\chi\propto \Theta_{\chi}(1+z_\chi)^2/H(z_\chi)$ and $P_\chi(T_\chi)\propto H(z_\chi)/[\Theta_\chi(1+z_\chi)^{2}]$.  Hence both argument and function depend on the details of the dark matter. Therefore, for a given $\nu_{\rm obs}$ and $\Theta_\chi$, the heavier the dark matter, the higher the brightness temperature associated to its decays. Moreover, since at higher redshifts $\mathcal{P}_{\breve{\rho}}$ tends to a narrower distribution centered at $\breve{\rho}_{\rm m}=1$ due to smaller clustering, it may have a stronger effect on the total VID. Note also that for low enough $\Theta_{\chi}$, the brightness temperature due to dark-matter decays would be significantly smaller than the one associated with atomic or molecular lines, and the effect on the total VID would be minimal.

\begin{figure}[t]
 \begin{centering}
\includegraphics[width=\columnwidth]{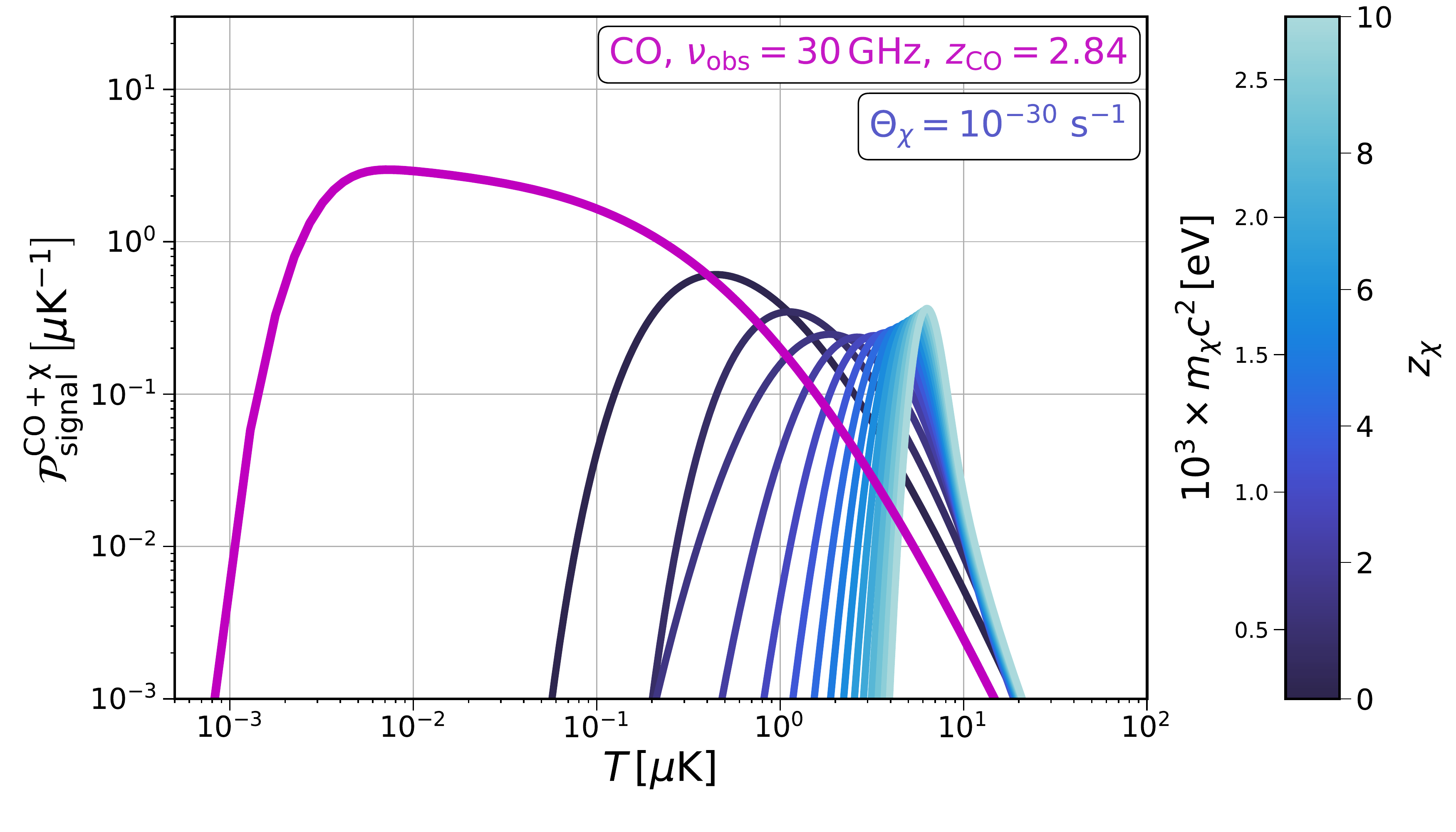}
\caption{Voxel intensity distribution of the CO spectral line (magenta) and the total emission including contributions from dark-matter decay with $\Theta_\chi=10^{-30}$ s$^{-1}$ for an experiment like COMAP 2 (without including instrumental noise). We consider dark-matter masses corresponding to $z_\chi\in [0,10]$.}  
\label{fig:VIDtot}
\end{centering}
\end{figure}

We illustrate the effect that an extra contribution to the observed brightness temperature from decaying dark matter  has in the VID of the CO line in Fig.~\ref{fig:VIDtot}. Note that the shift of the VID towards higher values of T is influenced by $\Theta_\chi$, but the shape of the overall VID depends on $z_\chi$, and thus $m_\chi$. Therefore, we expect that the mass of the dark matter can also be determined in a VID search. As was the case for the power-spectrum searches, VID analyses are limited by instrumental noise, leaving space for improving the sensitivity to dark-matter decays. However, unlike the case of the power spectrum, we expect a more significant signature from the dark-matter contribution for higher values of $m_\chi$.

Fortunately, since the VID is an one-point statistic and the probability distribution function associated to dark-matter decays is very different to those for atomic and molecular lines, there is no complication with the correlation between the targeted and the interloper lines. Therefore, we do not need to remove  the dark-matter masses corresponding to the redshifts probed by the targeted line from our analysis. 

\renewcommand{\arraystretch}{1.5}
\begin{table*}[t!]
\centering
\resizebox{\textwidth}{!}{%
\begin{tabular}{|c|c|c|c|c|c|c|c|c|c|c|c|}
\hline
Experiment & Line & $\nu_{\rm obs}$ {[}GHz{]} & $\Delta\nu$ {[}MHz{]} & $\delta\nu$ {[}kHz{]} & $T_{\rm sys}$ {[}K{]} & $D_{\rm dish}$ {[}m{]} & $D_{\rm min}$ {[}m{]} & $D_{\rm max}$ {[}m{]} & $\Omega_{\rm field}$ {[}deg$^2${]} & $t_{\rm obs}$ {[}hours{]} & $N_{\rm tot}$ \\ \hline\hline
HERA & HI & 0.11,0.13,0.15,0.18 & 18.9, 22.5, 26.8, 31.8 & 97.8 & $100+120\left(\frac{\nu_{\rm obs}}{150\, {\rm MHz}}\right)^{-2.55}$ & 14 & 14.6 & 876 & 1440 & 3000 & 700 \\ \hline
CHIME & HI & 0.44, 0.52, 0.62, 0.74 & 76, 90, 107, 127 & 390 & 50 & 20 $^a$ & 0.4 & 60 & 15000 $^b$ & 10000 & 2048 \\ \hline
HIRAX $^c$ & HI & 0.44, 0.52, 0.62, 0.74 & 76, 90, 107, 127 & 390 & 50 & 6 & 2 & 250 & 15000 & 10000 & 2048 \\ \hline\hline
COMAP 1 & CO & 30 & 8000 & 7324 & 40 & 10.4 & - & - & 2.25 & 6000 & 19 \\ \hline
COMAP 2 & CO & 30 & 8000 & 7324 & 40 & 10.4 & - & - & 60 & 10000 & 95 \\ \hline
\end{tabular}%
}
\caption{Specifications for experiments observing at centimeter wavelengths, which use the conventions of brightness temperature. Note that HERA, CHIME and HIRAX are interferometers, while COMAP is a dish-autocorrelation only experiment, and that we use $N_{\rm tot}=N_{\rm ant}N_{\rm feeds}N_{\rm pol}$.\\
$^a$ CHIME employs four contiguous cylinders of $20\times 100$ meters with 256 double-polarization detectors each. 20 m is the equivalent $D_{\rm dish}$ value to compute $\Omega_{\rm FoV}$.\\
$^b$ CHIME has the potential to scan half of the sky every day. However, we limit $\Omega_{\rm field}$ to account for masks covering the Milky Way and ensure that the footprint of CHIME and HIRAX do not overlap.\\
$^c$ FOR VID analyses of the two lowest redshift bins of HIRAX, we combine each two frequency channels, as well as use combined pixels of 2x2 pixels for the lowest redshift bin, to have larger voxels and ease the computations.}
\label{tab:cmexps}
\end{table*}
\renewcommand{\arraystretch}{1}

\begin{table*}[]
\centering
\resizebox{\textwidth}{!}{%
\begin{tabular}{|c|c|c|c|c|c|c|c|}
\hline
Experiment & Line & $\nu_{\rm obs}$ {[}GHz{]} & $\Delta\nu$ {[}GHz{]} & R & $\sigma_{\rm FWHM}\, ^a$ {[}arcsec{]} & $\Omega_{\rm field}$ {[}deg$^2${]} & $\sigma_{{\rm N},I}$ {[}Jy/sr{]} \\ \hline\hline
CCAT-prime & CII & 220, 280, 350, 408 & 40, 40, 40, 40 & 100 & 57, 45, 35, 30 & 8 & (0.6, 1.0, 2.5, 5.7)$10^4$ \\ \hline
AtLAST & CII & 345, 422, 543, 760 & 63, 95, 158, 317 & 1000 & 4.4, 3.6, 2.8, 2.0 & 7500 & (0.4, 0.7, 1.4, 3.9)$10^5$ \\ \hline\hline
SPHEREx & H$\alpha$ & (83, 109, 157, 295)10$^3$ & (20, 32, 65, 210)10$^3$ & 41.4 & 6.2 & 200 & 733, 747, 948, 1030 \\ \hline\hline
SPHEREx & Ly-$\alpha$ & (252, 308, 366)10$^3$ & (63, 48, 68)10$^3$ & 41.4 & 6.2 & 200 & 1006, 1022, 981 \\ \hline
HETDEX & Ly-$\alpha$ & (578, 647, 724, 811)10$^3$ & (65, 73, 82, 92)10$^3$ & 700 & 5.47 & 300+150$^b$ & 57 \\ \hline 
\end{tabular}%
}
\caption{Specifications for experiments observing at sub-centimeter wavelengths, which use the conventions of specific intensity.  \\
$^a$ For VID analyses we combine the pixels of some experiments to have larger voxels and ease the computations. We consider combined pixels of $30\times 30$, $2\times 2$, and $5\times 5$ for AtLAST, SPHEREx and HETDEX, respectively.\\
$^b$ HETDEX will observe two different fields, one on the spring sky and another on the fall sky, covering 300 and 150 deg$^2$, respectively. We assume that the observing time is distributed in a way that $t_{\rm pix}$ is the same for both fields.}
\label{tab:subcmexps}
\end{table*}

\section{Detection prospects}
\label{sec:forecasts}
As shown in Fig.~\ref{fig:landscape}, there are many upcoming LIM experiments, covering a vast frequency range, which translate to a wide mass range for decaying dark-matter particles. We aim to exploit this rich experimental landscape and consider HI, CO, CII, H$\alpha$, and Lyman-$\alpha$ lines. We assume our fiducial astrophysical models to follow the relation $L(M,z)$ between the line luminosity and the halo mass from Refs.~\cite{Padmanabhan_21cmLofM}, \cite{Li_CO_16}, \cite{Silva_CII}, \cite{Silva_Lyalpha} and \cite{Gong_lines} for post-reionization HI, CO, CII, Lyman-$\alpha$, and H$\alpha$ emission, respectively. We make an exception for the HI emission during ionization and model directly the mean temperature and the luminosity weighted bias, as done in e.g.,~\cite{SatoPolito_antisym}. We assume that the mean neutral hydrogen fraction evolves with redshift as~\cite{Pritchard_EoR,Kovetz_EDGES}
\begin{equation}
    \langle x_{\rm HI}\rangle (z) = \frac{1}{2}\left[1+\tanh \left(\frac{z-z_r}{\Delta z_r}\right)\right],
\end{equation}
where $z_r$ and $\Delta z_r$ are the midpoint and duration of reionization, respectively; we adopt $z_r=8$ and $\Delta z_r=0.5$ as our fiducial parameters. The mean brightness temperature is related to the mean neutral fraction as $\langle T\rangle \approx 27\langle x_{\rm HI}\rangle \sqrt{(1+z)/10}$ mK. Regarding the bias, we follow the parametrization 
\begin{equation}
    b^{\rm HI}(z) = \eta\left(\langle x_{\rm HI}\rangle(z)-1\right)+1,
\end{equation}
taking the value $\eta=14.8$ from fits to semi-numerical simulations~\cite{Hoffmann_biasEoR} as our fiducial value. 

We consider the following experiments: HERA~\cite{HERA}, HIRAX~\cite{HIRAX} and CHIME~\cite{CHIME} for HI (HERA probes the epoch of reionization and HIRAX and CHIME the post-reionization Universe), COMAP~\cite{Cleary_COMAP} (phases 1 and 2) for CO, CCAT-prime~\cite{CCAT-prime} and AtLAST~\cite{ATLAST} for CII, HETDEX~\cite{HETDEX} for Lyman-$\alpha$, and SPHEREx~\cite{SPHEREx} for Lyman-$\alpha$ and H$\alpha$. We combine results from HIRAX and CHIME (since they observe the south and north galactic caps, respectively), as well as SPHEREx H$\alpha$ and Lyman-$\alpha$ surveys, since they do not overlap in redshift. Details on the experimental specifications of all these experiments can be found in Tables~\ref{tab:cmexps} and~\ref{tab:subcmexps}.

We use the Fisher-matrix formalism~\cite{Fisher:1935,Jungman:1995av,Jungman:1995bz,Tegmark_fisher97} to forecast the sensitivity of power-spectrum measurements and the VID to decaying dark matter. Note that, since we do not model the covariance between the LIM power spectrum multipoles and the VID, we only consider these probes separately. We assume a $\Lambda$CDM Universe as our fiducial case, which corresponds to $\Theta_\chi=0$. In all cases all redshift bins and fields observed by each experiment are independent, and therefore the only parameter common to all of them is $\Theta_\chi$. 

\subsection{Power spectrum}
\label{sec:Pklim}

\begin{figure}[t]
 \begin{centering}
\includegraphics[width=\columnwidth]{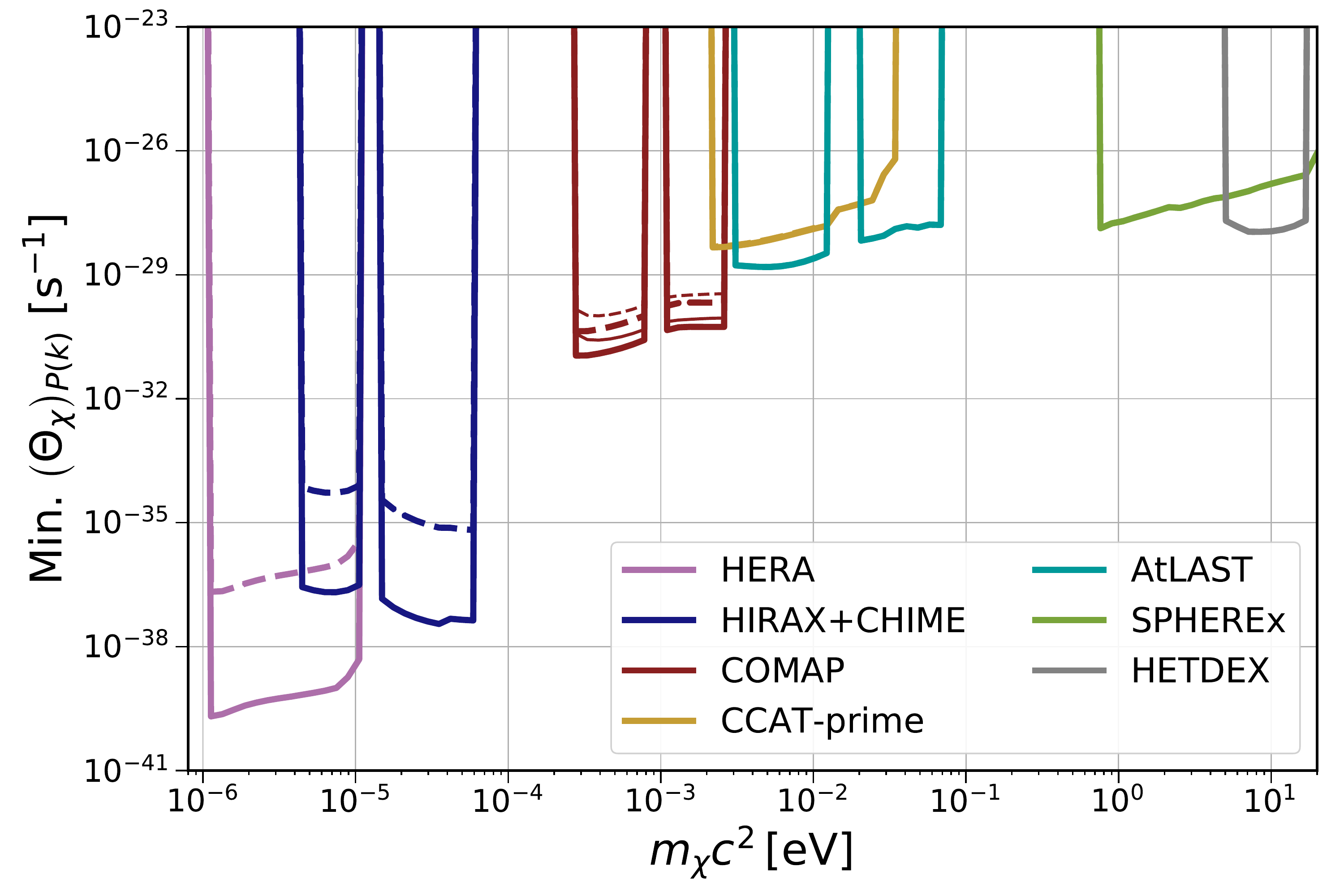}
\caption{Minimum $\Theta_\chi$ values that could be detected at 95\% confidence level as function of fixed dark-matter particle masses using measurements of the power spectrum multipoles of each LIM survey considered. Dashed lines do not include the contribution from the stimulated emission to ease the interpretation of these results for other models of radiative decaying dark matter. Thin and wide red lines correspond to COMAP 1 and COMAP 2, respectively.}  
\label{fig:sigmaPk}
\end{centering}
\end{figure}

We use the quantity $\Theta_\chi^2$ to parametrize the amplitude of the dark-matter contribution to the power spectrum, and follow Ref.~\cite{Li_DMdisentangle} to obtain upper limits to $\Theta_\chi$. We include the phenomenological parameters $\lbrace \left( \langle T\rangle b \right)^2, P_{\rm shot},\sigma_{\rm v}^2,\rbrace$ in our Fisher analysis and then marginalize over them for each redshift bin and field observed.   The parameter $\sigma_{\rm v,\chi}^2$ controlling small-scale redshift-space distortions at $z_\chi$ might also be included in the Fisher matrix analysis. However, we decide to leave it to future work because its effect is significantly smaller than that of $\Theta_\chi$, and it will only be important for a scenario where the radiation from dark-matter decays is detected.

We compute the Fisher matrix for each redshift bin and field observed, and marginalize over all parameters but $\Theta_\chi^2$. Afterwards, we generate numerical samples of $\Theta_\chi^2$ assuming a Gaussian distribution with the inverse of the sum of the marginalized Fisher matrices as variance. Finally, we apply a flat prior in $\Theta_\chi$, enforce $\Theta_\chi^2>0$, transform the samples to $\Theta_\chi$, and estimate the corresponding sensitivity as the minimum value of $\Theta_\chi$ that could be measured for a specific $m_\chi$ at 95\% confidence level.

We perform this process for all dark-matter particle masses between $10^{-6}$ and $20$ eV, for which $z_\chi\in \left[0.05, z_{\rm min}\right]\cup \left[z_{\rm max}, 10\right]$, where $z_{\rm min}$ and $z_{\rm max}$ are the limiting redshifts of the observed redshift bin.  We use $k_{\rm max}=0.5\, {\rm Mpc}^{-1}$ as the maximum value of $k$ in the forecast. We model the power-spectrum covariance and window functions as described in Sec.~\ref{sec:LIMPk} and App.~\ref{app:LIMPk}. We  neglect foregrounds for molecular lines, but model the loss of information due to foreground removal for HI surveys. We amplify the restriction of the volume window function $W_{\rm vol}$, assuming $N_\perp=N_\parallel=2$ for post-reionization HI observations (HIRAX and CHIME)~\cite{Soares_HImultipoles}. For HI observations during the epoch of reionization (HERA) we implement a foreground wedge, for which all modes $k_\parallel < k_{\parallel, {\rm min}} = A + Bk_\perp$ are lost~\cite{Pober_wedge13,Pober_wedge14}. We adopt $A=0.05$ $h/{\rm Mpc}^{-1}$ ($\sim 0.034$ Mpc$^{-1}$ for our fiducial choice of $h$), and $B=6$, values that correspond to a standard moderate scenario of foreground removal for HERA.

We show forecasted minimum values of $\Theta_\chi$ for the decay into two photons that could be detected at the 95\% confidence level using the LIM power spectrum multipoles to be measured by the experiments considered in this work  in Fig.~\ref{fig:sigmaPk}. HI surveys like the combination of HIRAX and CHIME and, particularly, HERA (because it targets HI emission at higher redshift), are the most sensitive ones. As the frequency bands of the experiments shift towards higher frequencies (probing heavier dark-matter masses), the forecasted  sensitivity become weaker. This, which is consistent with the results of Ref.~\cite{Creque-Sarbinowski_LIMDM}, is due to lower values of $X_{\rm LT}$ and $\mathcal{F}_\gamma^{\rm CMB}$, so that $P_\chi$ is smaller for the same $\Theta_\chi$. 

\subsection{VID}

\label{sec:VIDlim}
\begin{figure}[t]
 \begin{centering}
\includegraphics[width=\columnwidth]{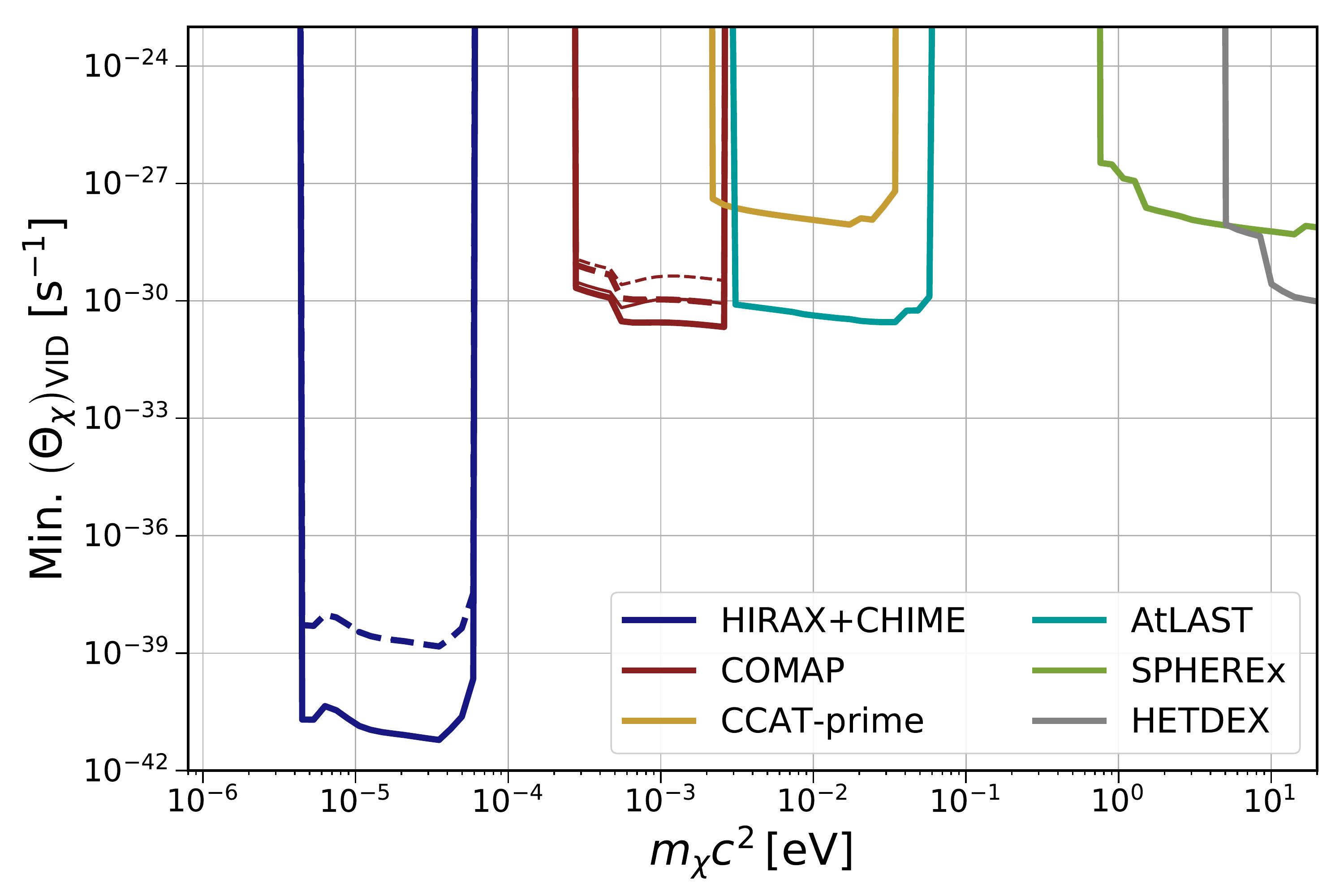}
\caption{Forecasted 95\% confidence level marginalized upper limits of $\Theta_\chi$ as function of the dark-matter particle mass from measurements of the VID of each LIM survey considered. Dashed lines do not include the contribution from the stimulated emission to ease the interpretation of these results for other models of radiative decaying dark matter.  Thin and wide red lines correspond to COMAP 1 and COMAP 2, respectively.}  
\label{fig:sigmaVID}
\end{centering}
\end{figure}

Each of the astrophysical models considered for each atomic or molecular line included in this analysis involves different assumptions and parameters. Nonetheless, the important quantity for VID studies is the luminosity function, rather than $L(M,z)$. Therefore, in order to homogenize the Fisher matrix analysis for VID measurements, we consider a modified Schechter function as the luminosity function~\cite{Schechter_function}:
\begin{equation}
    \frac{{\rm d}n}{{\rm d}L}
    = \phi_\star\left(\frac{L}{L_\star}\right)^\alpha\exp\left\lbrace -\frac{L}{L_\star}-\frac{L_{\rm min}}{L}\right\rbrace,
    \label{eq:SchCut}
\end{equation}
where $\phi_\star$, $L_\star$, $L_{\rm min}$ and $\alpha$ are free parameters parameters. We obtain the luminosity function for each line at each redshift of interest from $L(M,z)$ and the halo mass function as detailed in App.~\ref{app:dndL}, and use it to find the best-fit values of the free parameters of Eq.~\eqref{eq:SchCut}. We use these values, reported for each case in Table~\ref{tab:SchCut_params}, as the fiducial values for our Fisher matrix analysis.

Taking this into account, the parameters we vary in the Fisher matrix analysis of each redshift bin and field observed are $\lbrace{\phi_\star,\, L_\star,\, L_{\rm min},\, \alpha,\, \Theta_\chi \rbrace}$. As in the case for the power spectrum, we need to apply a prior enforcing $\Theta_\chi>0$. Hence, we proceed similarly but applying only this prior and working directly with $\Theta_\chi$.  Contrarily to the power spectrum, we use all dark-matter masses between $10^{-6}$ and $20$ eV for which $z_\chi\in \left[0.05, 10\right]$.

By default, we use the angular and spectral resolution of the experiment to define the voxel. However, if the resolution is too good, the resulting voxels would be too small and too many of them would not contain any emitter. Therefore in some cases we combine pixels or spectral channels to obtain larger voxels: we combine each two frequency channels for the two lower redshift bins of HIRAX (and use 2x2 pixels for its lowest redshift bin), and use combined pixels of 30x30, 2x2 and 5x5 for AtLAST, SPHEREx and HETDEX, respectively. 

Contrarily to experiments using only the auto-correlation of their antennas, interferometers are not well suited to measure the total brightness temperature, but its spatial fluctuations.\footnote{Interferometers are sensitive to the total mean brightness temperature only if certain conditions are fulfilled, see e.g.,~\cite{Presley_T0interf}.} Moreover, the presence of bright foregrounds for HI observations add  additional difficulties to the measurements. We avoid these complications considering the probability distribution function $\mathcal{P}_{\delta T} (\delta T) \equiv\mathcal{P}(\delta T+\langle T\rangle)$ of just the temperature fluctuations $\delta T$~\cite{Breysse_VID} for CHIME and HIRAX. Note that in this case the mean temperature also includes the noise root mean square. Finally, we do not include HERA in this analysis because we cannot compute the VID for the HI emission from the epoch of reionization with the model assumed. 

We show the forecasted minimum $\Theta_\chi$ values
that could be detected at the 95\% confidence level  as function of dark-matter mass using VID measurements from each LIM survey considered in this work in Fig.~\ref{fig:VIDtot}. While the general trend of having weaker upper limits for higher masses appears in this case too, we forecast higher sensitivity than using the power spectrum for most experiments. This is not surprising, since, for standard spectral lines, the VID is more sensitive to the luminosity function than the power spectrum (which is more sensitive to the cosmology). In turn, the contribution from dark-matter decays to the power spectrum that is not degenerate with the astrophysical uncertainties of the atomic or molecular line is mostly encoded in the quadrupole and hexadecapole, which have lower signal-to-noise ratio.

However, the largest difference with respect to the forecasts from power spectrum measurements is that in this case the sensitivity within the same experiment increases for more massive dark-matter particles. This is because for the same $\nu_{\rm obs}$, decays of more massive dark-matter particles happen at higher redshift, which correspond to higher brightness temperatures. Therefore, it is easier that the emission from dark-matter decays dominate the VID in this case than for lighter masses. As in the previous subsection, we also show results without the stimulated emission to ease the interpretation of the results for other models.

\section{Discussion}
\label{sec:discussion}
The results reported in the previous Section show the great promise of  LIM surveys to seek the product of radiative dark-matter decays. While we build upon the ideas discussed in Ref.~\cite{Creque-Sarbinowski_LIMDM}, here we propose more realistic strategies that are achievable and will not require further observations or dedicated searches. 

The main difference with respect to previous works is that we do not look just for an excess of radiation over the expected smooth  extra-galactic background light using the power spectrum. Instead, we account for the atomic and molecular spectral lines that will be targeted by LIM experiments and model the effect of unaccounted dark-matter decays on the LIM measurements. Applying methodologies designed to detect and account for line interlopers in LIM experiments, we interpret the unexpected contribution from dark-matter decays as another interloper, model its contribution, and turn it into our targeted signal. Furthermore, we perform a Fisher-matrix analysis to account for potential degeneracies with astrophysical uncertainties  of the standard LIM signal and obtain a more reliable estimation of the sensitivity to detect dark-matter decays. 

\begin{figure}[t]
 \begin{centering}
\includegraphics[width=\columnwidth]{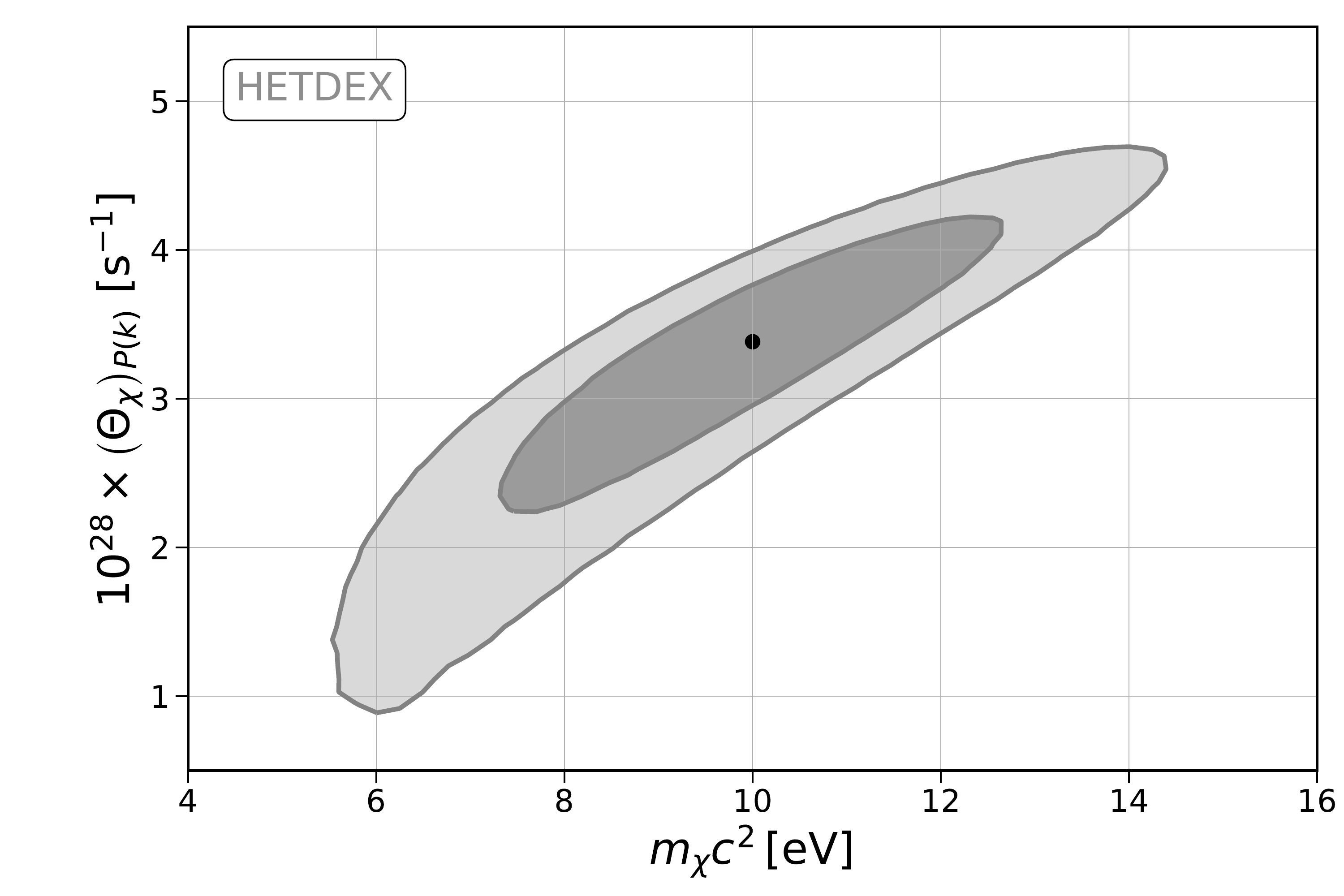}
\includegraphics[width=\columnwidth]{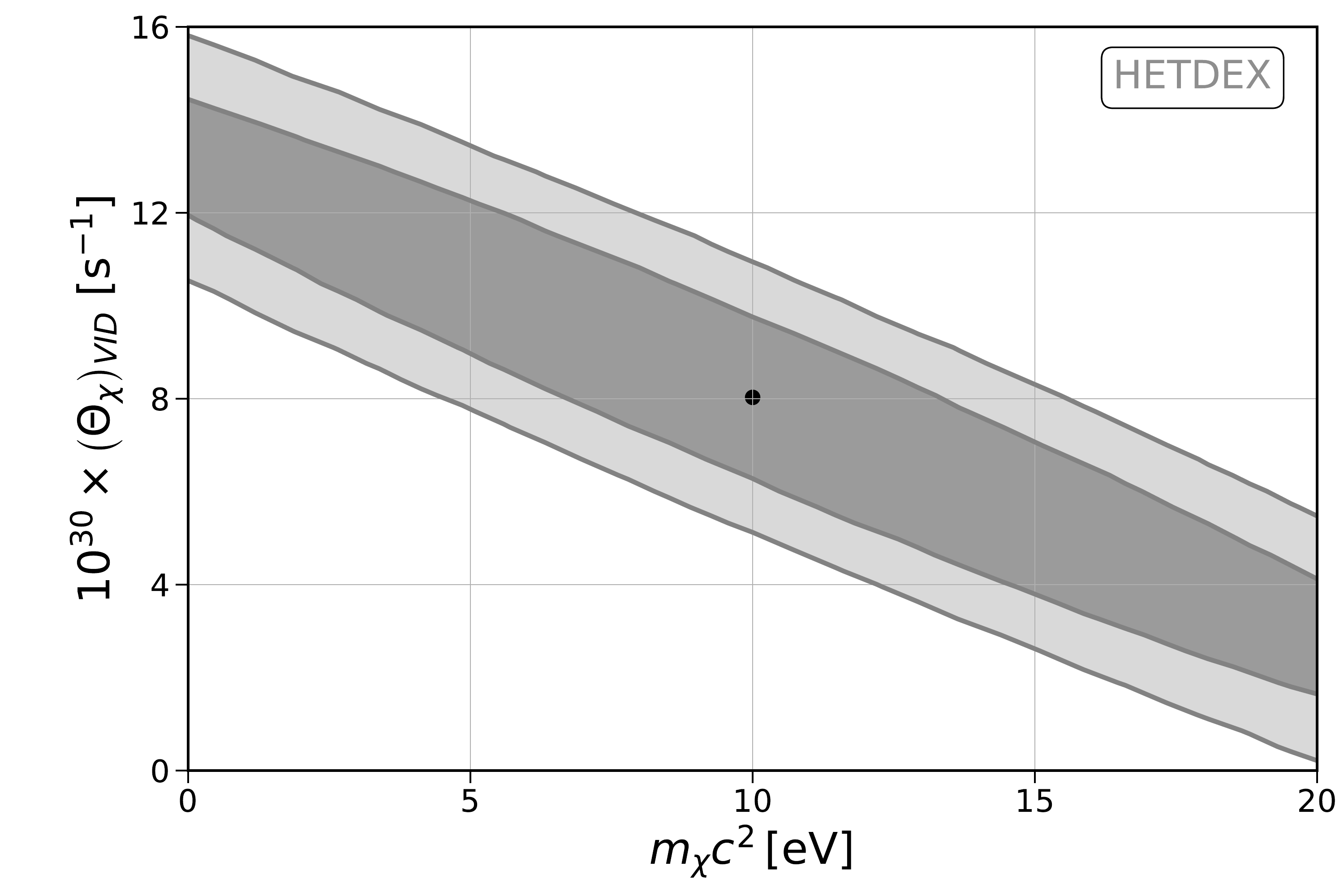}
\caption{68\% and 95\% confidence level marginalized constraints on the $\Theta_\chi-m_\chi c^2$ plane for an assumed decaying dark-matter model that would grant a detection by HETDEX using the power spectrum (top) and VID (bottom). We assume $m_\chi c^2=10$ eV and $\Theta_\chi$ to be three times the minimum values that would grant a 95\% confidence level detection. The fiducial model assumed is marked by a black dot. Note the change of scale in both axes between the two panels.}
\label{fig:2D}
\end{centering}
\end{figure}

The sensitivity estimates above (Figs.~\ref{fig:Pkproj} and~\ref{fig:VIDtot}) have assumed a search for a dark-matter particle of a specific mass.  If, however, that mass is unknown {\it a priori} and the search scans over all masses, then the interpretation of a ``3-sigma" detection would need to take into account the marginalization over the unknown dark-matter mass.
 
To assess the effects of marginalization over the dark-matter mass, we consider a decaying dark-matter model that, according to the previous Section, would be detected by LIM experiments. We assume a dark-matter mass of $10$ eV, and $\Theta_\chi$ values three times larger than the minimum values that would grant a 95\% confidence level detection by HETDEX, as reported in Figs.~\ref{fig:sigmaPk} and~\ref{fig:sigmaVID}. This results in $\Theta_\chi = 3.4\times 10^{-28}$ s$^{-1}$ and $8.0\times 10^{-30}$ s$^{-1}$ for the power spectrum and the VID, respectively. In this case, we repeat the  Fisher matrix analysis as described above, but add the dark-matter mass as a free parameter. Moreover, since we are assuming a detection, we also add $\sigma_{\rm v,\chi}^2$ in the forecast regarding the power spectrum.

We demonstrate the sensitivity of our proposed strategies to dark-matter masses in Fig.~\ref{fig:2D}, where we show the marginalized forecasted constraints to the $\Theta_\chi$-$m_\chi c^2$ plane for the dark-matter model mentioned above from HETDEX LIM measurements. We see how the dark-matter mass can be constrained with the LIM power spectrum. The dark matter mass and its decay rate are very degenerate, but this degeneracy can be broken thanks to the information from the LIM power spectrum.  

\subsection{Comparison to other limits for axion models}
Throughout this work we have considered radiative dark-matter decays without specifying a model, focusing on the decays into two photons to forecast the sensitivity of LIM experiments to detect these decays. Nonetheless, as explained in the end of Sec.~\ref{sec:DMDIM}, our results are general and can be recast to any dark-matter model that decays into two particles, at least one of them being a photon.

\begin{figure}[t]
 \begin{centering}
\includegraphics[width=\columnwidth]{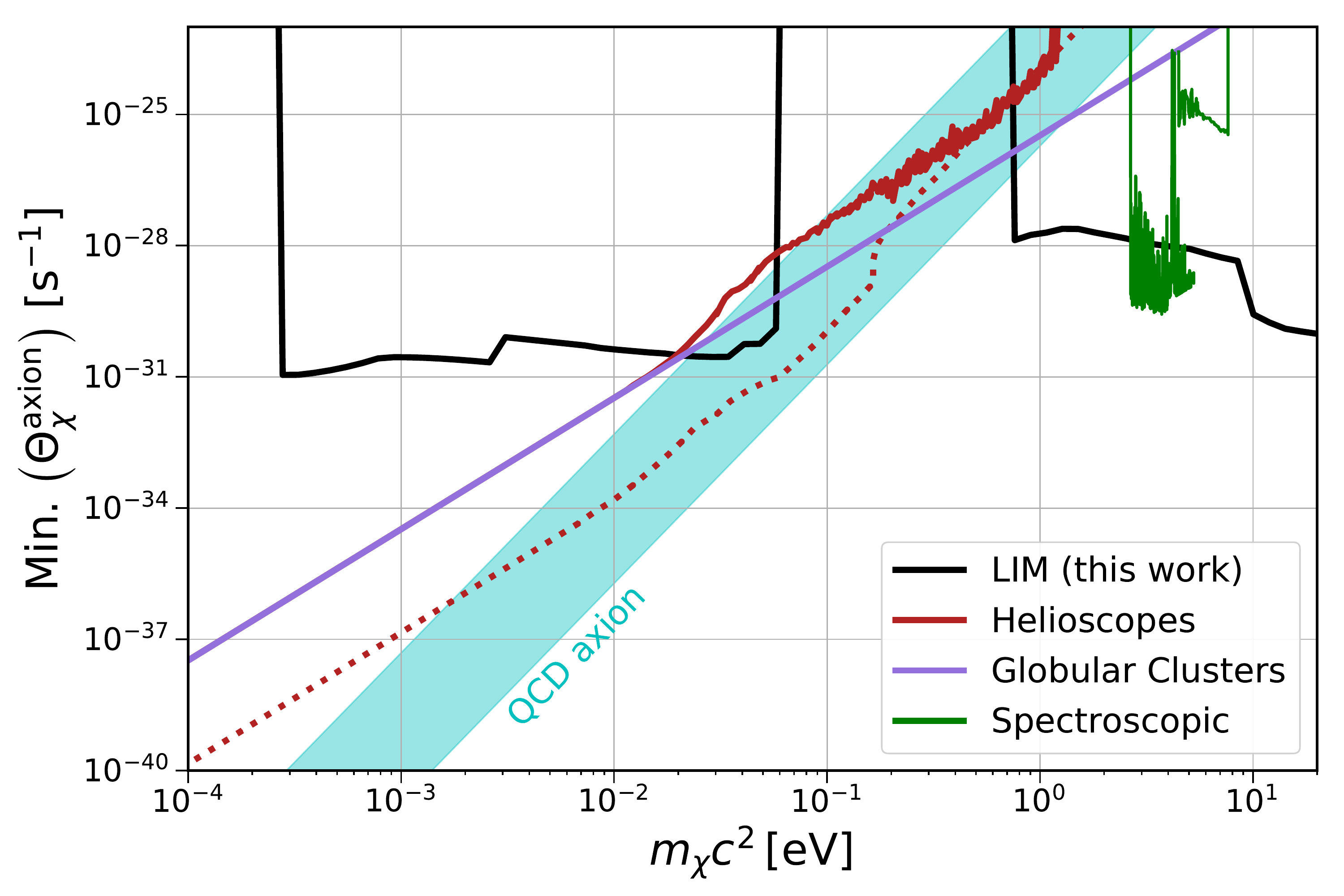}
\caption{Comparison between the minimum values of $\Theta_\chi$ that would be detectable by LIM surveys at 95\% confidence level derived in this work (choosing, but not combining, between results from the LIM power spectrum and the VID) interpreted in the context of axion dark matter and existing bounds and forecasted sensitivity (dotted lines) from independent observables, at the same significance. We include the theoretical prediction for the QCD axion in cyan.}
\label{fig:axions}
\end{centering}
\end{figure}

Still, the axion is a longstanding and compelling dark-matter candidate that decays to two photons and that has been sought through a variety of experimental and observational avenues. We compare the forecasted sensitivity of LIM experiments to detect axions with existing and forecasted bounds in Fig.~\ref{fig:axions}. We see that LIM experiments have the potential to significantly contribute to the search for axions, especially for masses $\sim 1-10$ eV. This mass range is  difficult to probe with helioscopes due to loss of coherence in the axion-photon conversion probability. Therefore, the most competitive bounds in this mass range correspond to spectroscopic observations of the dwarf spheroidal galaxy Leo T with MUSE~\cite{Regis:2020fhw} and searches for optical line emission in the galaxy clusters Abell 2667 and 2390, using VIMOS spectra~\cite{Grin:2006aw} (both denoted as `spectroscopic' in the figure). SPHEREx and HETDEX have the potential to improve the sensitivity to axion decays several orders of magnitude compared to existing and forecasted bounds from other probes.  
In addition AtLAST will probe the QCD axion for masses $\sim 10^{-2}-10^{-1}$, being more sensitive than current bounds from the helioscope CAST~\cite{Anastassopoulos:2017ftl} and the cooling of horizontal branch stars in globular clusters~\cite{Ayala:2014pea}, 
reaching a potential comparable to the next-generation helioscope IAXO~\cite{Armengaud:2019uso}. 

Note that the bounds using VIMOS and MUSE observations shown in~Fig.~\ref{fig:axions} essentially assume $f_{\chi} f_{\rm esc} f_{\gamma\gamma} = 1$, while bounds from helioscopes and globular clusters do not assume the axion to be the dark matter. Therefore, $\Theta_\chi = \Gamma_\chi$ for all these results. Nonetheless, for axion (like) particles we can assume $f_{\gamma\gamma} \simeq 1$. In addition, if we assume that all the dark matter is made of axions, we can assume $f_\chi  = 1$, and we expect $f_{\rm esc}$ close to $1$ for the masses considered in this work.  Taking this into account, the comparison between our results and other axion searches in Fig.~\ref{fig:axions} is appropriate.

\subsection{Extending and improving dark-matter detection with LIM}
We have focused on the LIM power spectrum and VID in this work. However, the basic concept of our strategies, namely interpreting the radiative dark-matter decays as an interloper of atomic or molecular lines, can be easily extended to other summary statistics of the line-intensity maps. The combination of different summary statistics will improve the sensitivity. While we do not combine the power spectrum and VID results because we do not model their covariance, we note that the correlation between $\Theta_\chi$ and $m_\chi$ is very different for each of them (see Fig.~\ref{fig:2D}). In addition, while the power spectrum is more sensitive to decays at lower redshift than the targeted emission, it is opposite for the VID. All this makes us expect large improvements upon consistent combination of different measurements. 

On the other hand, while we rely only on LIM observations of one line at a time, radiative decays can also be searched cross-correlating different spectral lines, or a line intensity and another tracer of the large scale structure such as galaxies.  The multi-tracer approach partially cancels the sample variance, boosting the significance of power spectrum measurements. This would boost the signal-to-noise ratio of the power spectrum quadrupole and hexadecapole, therefore improving the determination of the anisotropy of the power spectrum and the sensitivity to the dark-matter decay rate and mass. Furthermore, cross-correlating line-intensity fluctuations at different $\nu_{\rm obs}$ would probe two dark-matter masses at the same time. On the other hand, cross-correlations of LIM and galaxy surveys will have a smaller contribution from dark-matter decays.

The astrophysical modeling of the line emission is very uncertain, and strong variations from the expected model might be confused with the contribution from dark matter decays. On top of it, the effect of unaccounted for foregrounds may hinder the measurements. However, the strategies discussed above are robust against these uncertainties, and we have modeled the loss of information due to foreground contamination. The phenomenological parametrization of the astrophysical terms in the power spectrum allows us to marginalize over the astrophysical uncertainties without relying on a specific model. Since the LIM power spectrum is mostly sensitive to cosmology, a different astrophysical model for the emission would change the amplitude of the power spectrum, which would only affect the over all signal-to-noise ratio of the power spectrum measurement. This would change the sensitivity to the contribution from dark matter decays, but would not allow for a biased determination. The only case that may be problematic is the HI emission from reionization, for which variations of the model such as the presence of extreme clumping may alter also the shape of the power spectrum. We have used a simplistic model as illustration, but a more comprehensive modeling and marginalization may be required when dealing with actual observations. The VID is more sensitive to the astrophysics, but we marginalize over all free parameters of the Schechter function assumed to model the luminosity function, so we cover a wide range of astrophysical scenarios.

Nonetheless, as we have shown in Secs.~\ref{sec:DMDIM} and~\ref{sec:strategies}, we can reliably model the contribution from dark matter decays to the power spectrum and the VID. More importantly, this contribution is very characteristic in the two summary statistics we consider. Therefore, the combination of the VID and power spectrum will not only allow to increase the sensitivity to the dark matter but, given that both measurements are affected by astrophysical uncertainties and foregrounds in different ways, it will significantly increase the reliability of an eventual detection.  Multi-tracer studies will also increase the robustness against potential systematic errors in the analysis or measurements.

Finally, there are other ways to increase the relative contribution from dark-matter decays to the total LIM measurements. For instance, most of the spectral lines considered in this work are sourced by processes triggered by star formation. Masking the brightest sources 
can help to reduce the standard astrophysical contribution to the line intensity. 
This procedure can be optimized if the brightest sources are identified using external observations~\cite{Visbal_mask, Silva_CII, Sun_foregrounds}.  
The information from the dark-matter decays is not significantly affected by the mask because their contribution is not related with star formation at all. A similar approach has been proposed to reduce the contamination of line-interlopers (see e.g., Ref.~\cite{Breysse_foregrounds}).

\subsection{Detecting annihilations and non-radiative processes}

The search techniques discussed here can also be applied, with minimal modifications, to seek lines from dark-matter annihilation. If the dark matter is cold and annihilates to a two body final state with at least one photon, then that annihilation will produce an emission line. Note, however, that if dark matter annihilates to two photons that can be observed by LIM experiments, it will mean that dark matter is warm. 

Like the lines from dark-matter decay, annihilation lines trace the distribution of dark matter. However, the annihilation rate is proportional to the square of the dark matter density, rather than than have a linear relation as is the case of the decay rate. Therefore,  annihilation lines will be boosted in dense halos and will provide a biased tracer of the mass distribution. 

Although the contribution from the fluctuation signal we seek to the LIM power spectrum is thus expected to be more significant relative to the decay signal, for the same mean line luminosity density, the effects of biasing on the clustering will need to be taken into account. The VID from annihilation will, however, have to be modeled and will experience larger theoretical uncertainties due to imprecise knowledge of the details of dark-matter distributions within halos.  The VID from annihilations is also likely to more closely resemble the VID from atomic and molecular lines and its contribution to the total VID will thus be more difficult to distinguish.

\section{Conclusions}
\label{sec:conclusions}
Characterizing the nature of dark matter at a microscopic level is one of the main goals of cosmology, and considerable efforts have been made to address this challenge. One possibility is that dark-matter decays into Standard Model particles; more specifically, those decays may involve photons. In the case of a two-body decay, the produced photons are monoenergetic, in such a way that they become a well-defined emission line in the electromagnetic spectrum.

Here we have proposed two new techniques to seek radiative dark-matter decays with LIM.  Our analysis and forecasts take into account astrophysical uncertainties and limitations and capabilities of realistic experiments.  These proposed techniques can be applied to LIM measurements that are already planned, and they do not require any further augmentations of these projects nor dedicated programs.  Even so, they have the potential to open up new regions of discovery space, including some of the parameter space for the longstanding Peccei-Quinn axion.

The core idea of our proposal is that the dark-matter emission line will appear as a line interloper of the atomic and molecular spectral lines targeted by LIM experiments. However, unlike standard interloper lines, we do not take the contribution from dark matter as a contaminant but as the signal we look for, marginalizing over the astrophysical uncertainties dominating the contribution of the standard spectral lines. 

Building upon this idea, we have applied methodologies designed to detect and model the contribution from known faint lines that will contaminate LIM observations, adapting them to our purposes. Specifically, we have focused on the LIM power spectrum multipoles and the VID. The contribution from dark-matter decays to the LIM power spectrum, coming from different redshifts than the standard emission, is subject to projection effects that heavily increases the anisotropy of the power spectrum, in addition to contribute to the monopole. On the other hand, the probability distribution function of the brightness temperature associated to dark-matter decays is expected to be very different than the one related to astrophysical lines. Hence, the total probability distribution function, inferred using the VID, is expected to change its shape and shift towards higher temperatures due to the additional, exotic contribution.

We have considered a very general dark-matter scenario in which a fraction $f_\chi$ of the dark-matter decays with a decay rate $\Gamma_\chi$, a fraction $f_{\gamma x}$ of such decays involves at least a photon as daughter particle (where $x$ can be another photon or a different particle), and a fraction $f_{\rm esc}$ of such photons reaches the LIM experiment. Under these considerations, LIM observables are sensitive to $\Theta_\chi=f_\chi f_{\gamma x} f_{\rm esc}\Gamma_\chi$. After modeling the contribution from dark-matter decays as a line interloper for the measurements under study, we have forecasted the sensitivity to this exotic contribution of ongoing and forthcoming LIM experiments, spanning several orders of magnitude in frequency. We have found that in general the VID will be more sensitive to dark-matter decays than the power spectrum, but both of them are also sensitive to the dark-matter mass. Therefore, if a detection of decaying dark matter with LIM surveys is accomplished, it will also return information about the dark-matter mass. We report our results in the context of dark matter decaying into two photons, but provide a simple way to reinterpret our results for different daughter particles. 

More importantly, the potential sensitivity of LIM experiments will be extremely competitive compared to other observational probes. As an illustration, we interpret our results in the context of axion dark matter, and compare our forecasts with existing and forecasted bounds from other probes. We find that HETDEX and SPHEREx will improve the sensitivity to radiative dark-matter decays several orders of magnitude for axion masses of $1-10$ eV, while AtLAST will improve current constraints and present comparable potential than IAXO at masses $10^{-2}-10^{-1}$ eV.

LIM surveys hold great promise to address astrophysical and cosmological questions. Thanks to the strategies proposed in this work, LIM experiments will be extremely sensitive to radiative dark-matter decays, even without a dedicated program focused on dark matter. We eagerly await for these surveys to complete their observations to tackle the challenge of characterizing the nature of dark matter.

\acknowledgments
The authors thank Ely D. Kovetz for useful comments on the manuscript. JLB is supported by the Allan C. and Dorothy H. Davis Fellowship. AC acknowledges support from the the Israel Science Foundation (Grant No. 1302/19), the US-Israeli BSF (grant 2018236) and the German Israeli GIF (grant I-2524-303.7). AC acknowledges hospitality of the Max Planck Institute of Physics in Munich and of JHU where this work started, as well as support of the Centre of Cosmological Studies and the Balzan Foundation Award.  This work was supported at Johns Hopkins by NSF Grant No.\ 1818899 and the Simons Foundation.

\appendix

\section{Details on the modeling of LIM observables}
\label{app:LIMPk}
In this appendix we complete the description of the modeling of the LIM power spectrum and its covariance, filling the details that were left out of the main text. 

\subsection{Specific intensity}
In this work, we have used the brightness temperature to quantify the integrated emission of a given spectral line. However, this convention is used only for experiments observing millimeter and larger wavelengths. For sub-millimiter and shorter wavelength experiments, it is customary to use the specific intensity $I$, given by
\begin{equation}
    I(z) = \frac{c}{4\pi\nu H(z)}\rho_{\rm L}(z) = X_{\rm LI}(z)\rho_{\rm L}(z)\,.
\end{equation}
Throughout this work we have used $T$ for conciseness, but all the expressions can be adapted to the specific intensity changing $X_{\rm LT}$ by $X_{\rm LI}$.

\subsection{LIM bias and redshift-space distortions}
The bias factor relating brightness temperature and matter density perturbations can be approximated at large scales as the luminosity-weighted linear halo bias:
\begin{equation}
    b(z) = \frac{\int{\rm d}ML(M,z)b_{\rm h}(M,z)\frac{{\rm d}n}{{\rm d}M}(M,z)}{\int {\rm d}ML(M,z)\frac{{\rm d}n}{{\rm d}M}(M,z)},
\end{equation}
where $b_{\rm h}$ is the halo bias and we assume the halo mass function and halo bias fitting function from Ref.~\cite{Tinker_hmf2010}. Brightness temperature maps are taken in redshift space, hence the measured power spectrum is affected by redshift-space distortions. We model this effect with the Kaiser effect~\cite{Kaiser1987} at large scales and a Lorentzian function that empirically reproduces the effects at small scales. The resulting redshift-space distortions factor is given by
\begin{equation}
    F_{\rm rsd}(k,\mu,z) = \left(1+ \frac{f(z)}{b(z)}\mu^2\right)\frac{1}{1+0.5\left(k\mu\sigma_{\rm v}\right)^2},
\label{eq:RSD}
\end{equation}
where $f(z)$ is the growth rate.

\subsection{Instrumental resolution and window functions}
The angular resolution is related with the width of the effective beam profile, $\sigma_{\rm beam}=\theta_{\rm FWHM}/\sqrt{8\log 2}$, where $\theta_{\rm FWHM}=1.22c/\nu_{\rm obs}D$ is the full-width half maximum of the telescope beam, and $D$ is the diameter $D_{\rm dish}$ of the dish or the maximum baseline  distance $D_{\rm max}$ whether only the auto-correlation of each antenna is exploited or interferometric techniques are used, respectively. On the other hand, the spectral resolution is determined by the width of the frequency channel, $\delta\nu$. The characteristic resolution limits along the line of sight and transverse directions are
\begin{equation}
    \sigma_\parallel = \frac{c\delta\nu(1+z)}{H(z)\nu_{\rm obs}}, \qquad \sigma_\perp = D_{\rm M}(z)\sigma_{\rm beam},
\end{equation}
Note that a physical broadening of the spectral line due to peculiar velocities or dispersion of the photon frequency also limits a precise determination of the line, which may limit the spectral resolution if too accused. The effect of the line broadening can be folded in $\sigma_\parallel$, using the maximum between the broadening and $\delta\nu$. With this in mind, the resolution window can be computed as~\cite{Li_CO_16}:
 \begin{equation}
     W_{\rm res}(k,\mu) = \exp\left\lbrace -k^2\left[\sigma_\perp^2(1-\mu^2)+\sigma_\parallel^2\mu^2    \right]\right\rbrace.
 \label{eq:Wk_res}
 \end{equation} 

The volume within a solid angle $\Omega_{\rm field}$ and a frequency band $\Delta \nu$ probed by a survey is
\begin{equation}
    V_{\rm field} = L_\perp^2L_\parallel=\left[ D_{\rm M}(z)^2(z)\Omega_{\rm field}  \right]\left[ \frac{c(1+z)^2\Delta\nu}{H(z)\nu} \right].
\label{eq:Vfield}
\end{equation}  
In the absence of complex observation footprints, $V_{\rm field}$ can be considered as the volume observed by a survey. This volume determines the characteristic largest scales that are available, corresponding to Fourier modes $k_\perp^{\rm min, dish}\equiv 2\pi/L_\perp $ and $k_\parallel^{\rm min}\equiv 2\pi /L_\parallel$. The window function that accounts for the loss of modes beyond these scales is
\begin{equation}
\begin{split}
    W_{\rm vol}(k,\mu) = & \left(1-\exp\left\lbrace -\left(\frac{k}{N_\perp k^{\rm min}_\perp}\right)^2\left(1-\mu^2 \right) \right\rbrace \right)\times \\
& \times \left(1-\exp\left\lbrace -\left(\frac{k}{N_\parallel k^{\rm min}_\parallel}\right)^2\mu^2 \right\rbrace \right), 
\end{split}
\label{eq:Wk_vol}
\end{equation}
where we have introduced $N_\perp$ and $N_\parallel$ to allow for additional loss of modes due to the presence of foregrounds~\cite{Cunnington_HIforegrounds,Soares_HImultipoles}. 

\subsection{Covariance of the LIM power spectrum}
The definition of the total noise variance per voxel per antenna depends on the convention used to quantify the intensity of the line, either using brightness temperature or specific intensity. For low-frequency experiments that use $T$, the total variance per voxel and antenna is
\begin{equation}
    \sigma_{{\rm N},T}^2 = \frac{T_{\rm sys}^2}{N_{\rm feeds}N_{\rm pol}\delta\nu t_{\rm pix}},
\end{equation}
where $T_{\rm sys}$ is the system temperature of the telescope, $N_{\rm feeds}$ is the number of detectors in each antenna, $N_{\rm pol}=1,2$ is the number of polarizations that the detector is able to measure, and $t_{\rm pix}\equiv t_{\rm obs}/N_{\rm pix}$ is the observing time per pixel. We assume that the total observing time $t_{\rm obs}$ is uniformly distributed among all pixels. 

 In turn, for high-frequency experiments that work with specific intensities, the total variance per voxel and antenna is
 \begin{equation}
     \sigma_{{\rm N}, I}^2=\frac{\sigma_{\rm pix}^2}{N_{\rm feeds}N_{\rm pol}t_{\rm pix}},
 \end{equation}
 where $\sigma_{\rm pix}$ is typically given in terms of a noise equivalent intensity (NEI) in units of Jy s$^{1/2}$/sr. 

Regarding the experiments using interferometric techniques, we consider always a constant number density of baselines, for which
\begin{equation}
    n_{\rm s} = \frac{c^2N_{\rm ant}\left(N_{\rm ant}-1\right)}{2\pi\nu_{\rm obs}^2\left(D_{\rm max}^2-D_{\rm min}^2\right)}\,
    \label{eq:ns}
\end{equation}
where $D_{\rm min}$ is the minimum baseline distance. 

Finally, the number of modes observed per $k$ and $\mu$ bin, with respective widths $\Delta k$ and $\Delta\mu$, is
\begin{equation}
    N_{\rm modes}=\frac{V_{\rm field}k^2\Delta k\Delta \mu}{8\pi^2}\,.
\end{equation}
 Note that the loss of modes due to instrument resolution, foregrounds or the size of the volume probed is not modeled in $N_{\rm modes}$ but in the window functions, to properly account for all the accessible information in the multipoles of the LIM power spectrum~\cite{Bernal_IM}.

\subsection{Probability of the number of emitters in a voxel}
We approximate the halo number-count distribution with a lognormal distribution~\cite{Coles_LogNormal}, and assume that the expectation value $\eta$ for the number $N_{\rm e}$ of emitters within a voxel depends on the lognormal matter density field in that point. Under these assumptions $N_{\rm e}$ is a Poisson draw from a distribution with mean $\eta$:
\begin{equation}
    \mathcal{P}_{\rm e}=\int{\rm d}\eta\mathcal{P}_{\rm LN}(\eta)\mathcal{P}_{\rm Poiss}(N_{\rm e},\eta),
\end{equation}
where $\mathcal{P}_{\rm LN}$ and $\mathcal{P}_{\rm Poiss}$ are a lognormal and a Poisson distribution, respectively. $\mathcal{P}_{\rm LN}$ can be expressed in terms of the Gaussian random variable $\delta_G$ with standard deviation $\sigma_G$ as~\cite{Kayo_PDF01}:
\begin{equation}
 \mathcal{P}_{\rm LN}(\eta) = \frac{1}{\eta\sqrt{2\pi\sigma_G^2}}\exp\left\lbrace\frac{-1}{2\sigma_G^2}\left[\log\frac{\eta}{\bar{N}}+\frac{\sigma_G^2}{2}\right]^2\right\rbrace\, ,
 \label{eq:P_LN}
\end{equation}
where $\bar{N}=\bar{n}V_{\rm vox}$, and we take $\sigma_G$ to be the root-mean-square of the linear density contrast of the emitters:
\begin{equation}
    \sigma_G=\int \frac{{\rm d}^3\boldsymbol{k}}{\left(2\pi\right)^3} \left\lvert W(\boldsymbol{k})\right\lvert^2 b^2P_{\rm m}(k),
\label{eq:sigmaG}
\end{equation}
where $W(\boldsymbol{k})$ is the Fourier transform of the voxel window function. We use $W_{\rm res}$ (Eq.~\eqref{eq:Wk_res}) as the voxel window function, multiplying $\sigma_\perp$ or $\sigma_\parallel$ by the corresponding integer in the case of combining pixels or frequency channels.  

\begin{table}[]
\centering
\resizebox{\columnwidth}{!}{%
\begin{tabular}{cccccc}
\hline
Line & $z$ & $\phi_\star$ {[}Mpc$^{-3}L_\odot^{-1}${]} & $L_\star$ {[}$L_\odot${]} & $\alpha$ & $L_{\rm min}$ {[}$L_\odot${]} \\ \hline\hline
\multirow{4}{*}{HI} & 0.93 & $8.44\times 10^{-16}$ & $9.95\times 10^6$ & $-2.03$ & 1.1 \\ \cline{2-6} 
 & 1.29 & $2.60\times 10^{-16}$ & $1.31\times 10^7$ & $-2.10$ & 1.7 \\ \cline{2-6} 
 & 1.73 & $5.32\times 10^{-17}$ & $1.51\times 10^7$ & $-2.22$ & 2.6 \\ \cline{2-6} 
 & 2.24 & $2.67\times 10^{-18}$ & $1.85\times 10^{7}$ & $-2.42$ & 4.4 \\ \hline\hline
CO & 2.84 & $1.60\times 10^{-9}$ & $3.98\times 10^5$ & $-1.88$ & 102 \\ \hline\hline
\multirow{8}{*}{CII} & 1.50 & $1.67\times 10^{-11}$ & $7.00\times 10^7$ & $-1.42$ & 0.09 \\ \cline{2-6} 
 & 2.50 & $4.37\times 10^{-12}$ & $2.04\times 10^8$ & $-1.44$ & 0.29 \\ \cline{2-6} 
 & 3.50 & $5.52\times 10^{-12}$ & $2.07\times 10^8$ & $-1.47$ & 2.2 \\ \cline{2-6} 
 & 3.65 & $5.51\times 10^{-12}$ & $2.01\times 10^8$ & $-1.48$ & 2.8 \\ \cline{2-6} 
 & 4.43 & $1.82\times 10^{-12}$ & $3.65\times 10^8$ & $-1.54$ & 11.5 \\ \cline{2-6} 
 & 4.50 & $1.54\times 10^{-12}$ & $3.93\times 10^8$ & $-1.54$ & 13.0 \\ \cline{2-6} 
 & 5.79 & $1.47\times 10^{-12}$ & $2.34\times 10^8$ & $-1.67$ & 80.5 \\ \cline{2-6} 
 & 7.64 & $6.50\times 10^{-13}$ & $1.83\times 10^8$ & $-1.78$ & 229 \\ \hline\hline
\multirow{4}{*}{H$\alpha$} & 0.55 & $1.66\times 10^{-11}$ & $9.93\times 10^7$ & $-1.40$ & 0.19 \\ \cline{2-6} 
 & 1.90 & $2.88\times 10^{-12}$ & $2.98\times 10^8$ & $-1.47$ & 1.36 \\ \cline{2-6} 
 & 3.20 & $3.13\times 10^{-12}$ & $2.64\times 10^8$ & $-1.55$ & 15.6 \\ \cline{2-6} 
 & 4.52 & $1.53\times 10^{-12}$ & $3.18\times 10^8$ & $-1.61$ & 39.2 \\ \hline\hline
\multirow{7}{*}{Ly-$\alpha$} & 2.00 & $1.58\times 10^{-11}$ & $1.08\times 10^8$ & $-1.56$ & 26.8 \\ \cline{2-6} 
 & 2.40 & $7.71\times 10^{-12}$ & $1.9\times 10^8$ & $-1.56$ & 35.8 \\ \cline{2-6} 
 & 2.81 & $3.37\times 10^{-12}$ & $3.6\times 10^8$ & $-1.56$ & 48.2 \\ \cline{2-6} 
 & 3.27 & $2.48\times 10^{-12}$ & $5.49\times 10^8$ & $-1.55$ & 66.1 \\ \cline{2-6} 
 & 5.74 & $4.77\times 10^{-14}$ & $6.70\times 10^9$ & $-1.63$ & 889 \\ \cline{2-6} 
 & 7.01 & $2.93\times 10^{-14}$ & $6.43\times 10^9$ & $-1.70$ & 2875 \\ \cline{2-6} 
 & 8.78 & $2.22\times 10^{-15}$ & $1.24\times 10^{10}$ & $-1.84$ & 9863 \\ \hline
\end{tabular}%
}
\caption{Parameters for the modified Schechter function for the luminosity function in Eq.~\eqref{eq:SchCut}, for all atomic and molecular lines and redshifts under consideration.}
\label{tab:SchCut_params}
\end{table}

\section{Parameters of the double-exponential PDF of matter fluctuations}
\label{app:PDFchi}
Here we report the fitting functions for the parameters $\varsigma$, $\rho_0$ and $\rho_1$ of the double-exponential probability density function of the matter fluctuations (see Eq.~\eqref{eq:doublePDF}), as proposed in Ref.~\cite{Klypin_PDF}:
\begin{equation}
\begin{split}
    \varsigma   = & -2 -\frac{0.05}{g(z)}\times \\
    & \times \left(1-2.4\sigma_{\rm m}^{0.05}\exp\left\lbrace-\left[\frac{4.7}{\sigma_{\rm m}g(z)}\right]^2\right\rbrace\right)\, ,
\end{split}
\end{equation}
where
\begin{equation}
    g(z) = 0.075+\frac{0.25}{(1+z)^5}\,;
\end{equation}
\begin{equation}
    \rho_0 = 0.048+\frac{0.77}{\sigma_{\rm m}}\,;
\end{equation}
\begin{equation}
    \rho_{1} = 4.7\sigma_{\rm m}^{1.9}\exp\left\lbrace -\frac{2}{\sigma_{\rm m}}\right\rbrace\,.
\end{equation}
Note that the redshift dependence is encoded in $g(z)$ and $\sigma_{\rm m}$.

\section{Luminosity functions}
\label{app:dndL}
As described above, we use the relation $L(M,z)$ between the luminosity of a given line and the halo mass to model the luminosity density of the line and other required quantities to compute the LIM observables, especially related to the power spectrum, except for the HI line from reionization. The computation of the VID requires instead the luminosity function ${\rm d}n/{\rm d} L$ (see Eq.~\eqref{eq:probT1}). One possible way to obtain the luminosity function consists of inverting the mass-luminosity relation (i.e., using $L(M,z)$ to obtain $M(L,z)$) and combined it with the halo mass function. However, this is only possible when $L(M,z)$ is monotonous. 

To overcome this complication, we use the conditional probability distribution function $\mathcal{P}(L\lvert M)$ of having a luminosity $L$ coming from a halo with mass $M$. We assume a lognormal distribution with mean $L(M,z)$ for $\mathcal{P}(L\lvert M)$, which allows to introduce scatter in $L(M,z)$ that preserves the mean luminosity if needed, or a Dirac delta distribution $\delta_D(L-L(M,z))$ otherwise; we assume a scatter of 0.2 in $\log_{\rm 10}=0.2$ for all lines but CO, for which we assume a scatter of 0.3~\cite{Li_CO_16}. Thus, the luminosity function is given by
\begin{equation}
    \frac{{\rm d} n}{{\rm d} L} = \int {\rm d} M \mathcal{P}(L\lvert M)\frac{{\rm d} n}{{\rm d} M}\, .
    \label{eq:dndL}
\end{equation}

In order to homogenize the analysis, we fit the luminosity functions for each astrophysical model, spectral line and redshift under consideration to a modified Schechter function (see Eq.~\eqref{eq:SchCut}). We provide the corresponding parameters for each luminosity distance in Table~\ref{tab:SchCut_params}.

\bibliography{Refs.bib}
\bibliographystyle{utcaps}
\bigskip

\end{document}